\documentclass[iop]{emulateapj}
\usepackage{graphicx}
\usepackage{amsmath,amssymb}
\usepackage{mathtools}
\usepackage{textcmds}
\usepackage{txfonts}
\usepackage{natbib}
\usepackage{multirow}
\usepackage{color}
\usepackage{xspace}
\usepackage{ulem}
\usepackage[colorlinks=true,citecolor=blue]{hyperref}

\begin{document}
   \title{APODIZED PUPIL LYOT CORONAGRAPHS FOR ARBITRARY APERTURES. IV.\\REDUCED INNER WORKING ANGLE AND INCREASED ROBUSTNESS TO LOW-ORDER ABERRATIONS}

   \author{Mamadou N'Diaye, Laurent Pueyo, R\'emi Soummer}
   \affil{Space Telescope Science Institute, 3700 San Martin Drive, 21218 Baltimore MD, USA.}

  \begin{abstract}
   The Apodized Pupil Lyot Coronagraph (APLC) is a diffraction suppression system installed in the recently deployed instruments Palomar/P1640, Gemini/GPI, and VLT/SPHERE to allow direct imaging and spectroscopy of circumstellar environments. Using a prolate apodization, the current implementations offer raw contrasts down to $10^{-7}$ at 0.2\,arcsec from a star over a wide bandpass (20\%), in the presence of central obstruction and struts, enabling the study of young or massive gaseous planets. Observations of older or lighter companions at smaller separations would require improvements in terms of inner working angle (IWA) and contrast, but the methods originally used for these designs were not able to fully explore the parameter space. We here propose a novel approach to improve the APLC performance. Our method relies on the linear properties of the coronagraphic electric field with the apodization at any wavelength to develop numerical solutions producing coronagraphic star images with high-contrast region in broadband light. We explore the parameter space by considering different aperture geometries, contrast levels, dark-zone sizes, bandpasses, and focal plane mask sizes. We present an application of these solutions to the case of Gemini/GPI with a design delivering a $10^{-8}$ raw contrast at 0.19\,arcsec and offering a significantly reduced sensitivity to low-order aberrations compared to the current implementation. Optimal solutions have also been found to reach $10^{-10}$ contrast in broadband light regardless of the telescope aperture shape, with effective IWA in the $2-3.5\lambda/D$ range, therefore making the APLC a suitable option for the future exoplanet direct imagers on the ground or in space.
    \end{abstract}
 

   \shorttitle{Low-order aberration insensitive APLC solutions}

%

\section{Introduction}\label{sec:intro}
The directly imaged exoplanets so far \citep[e.g.][]{2008Sci...322.1345K,2008Sci...322.1348M, 2010Natur.468.1080M, 2009A&A...493L..21L, 2013ApJ...776L..17R,2013ApJ...774...11K} with the existing facilities have been analyzed photometrically, providing hints on their physical properties \citep{2011ApJ...730L..21H,2011ApJ...741...55S,2013ApJ...775...56K,2013ApJ...779L..26R,2014A&A...565L...4G}. Some of these detected companions have recently been the subject of first spectroscopic observations, giving access to chemical features of their atmosphere \citep{2013ApJ...768...24O,2013arXiv1307.1404L,2013ApJ...776...15C,2013Sci...339.1398K,2013arXiv1308.3859B,2013ApJ...779..153H}.

The number of exoplanets with direct images and spectral characterization is expected to significantly increase over the next few years with the recently deployed instruments on ground-based observatories Palomar/P1640 \citep{2011PASP..123...74H}, Subaru/SCExAO \citep{2010SPIE.7736E..71G}, Gemini/GPI \citep{2008SPIE.7015E..31M}, and VLT/SPHERE \citep{2008SPIE.7014E..41B}, sheding light on planet diversity, formation and evolution through the refinement of current models \citep{2009ARA&A..47..253O}. These exoplanet direct imagers will observe young/massive, self-luminous gaseous planets around nearby stars, relying on extreme adaptive optics capabilities for high-angular resolution images \citep[ExAO,][]{2005JOSAA..22.1515P,2006OExpr..14.7515F}, optimized coronagraphs with raw contrast down to $10^{-7}$ at inner working angle (IWA) for starlight diffraction suppression \citep{2000PASP..112.1479R,2003A&A...404..379G,2011ApJ...729..144S}, wavefront error calibration strategies for quasi-static speckles correction \citep{2007JOSAA..24.2334S,2010SPIE.7736E.179W,2010SPIE.7736E..77P}, and integral field unit for spectrum extraction of the detected companions. 
 
In the next decade, new high-contrast imaging instruments are envisioned to observe fainter companions down to old/light gaseous planets seen in reflected light and massive rocky planets: on the ground with Extremely Large Telescopes (ELTs) and dedicated instruments \citep[e.g. E-ELT/PCS,][]{2013aoel.confE...8K} or in space with missions such as WFIRST-AFTA \citep{2013arXiv1305.5422S}, probe-scale missions \citep{2010SPIE.7731E..67T,2010SPIE.7731E..68G,2013EPJWC..4715004B}, or the post-JWST flagship project ATLAST \citep{2012OptEn..51a1007P}. In addition to exquisite wavefront sensing methods \citep[e.g. Zernike wavefront sensors,][]{2011SPIE.8126E..11W,2013A&A...555A..94N} and control strategies \citep[e.g. deformable mirrors based methods,][]{2006ApOpt..45.5143S,2007ApJ...666..609P,Pueyo:09}, these future instruments will rely on novel high-performance coronagraphs to offer broadband raw contrasts down to $10^{-10}$, enabling spectroscopy of these planetary companions.

In the past two decades, many coronagraphic concepts have been proposed to remove starlight diffraction over a wide spectral band (see e.g. reviews in \citet{2006ApJS..167...81G} or \citet{2012SPIE.8442E..04M}). Most of these systems have shown capabilities to reach contrasts below $10^{-7}$ at IWA smaller than 4\,$\lambda_0/D$ over a finite bandwidth $\Delta\lambda$, where $\lambda_0$ and $D$ denote the central wavelength of the bandpass and the telescope diameter \citep{2012SPIE.8442E..04M}. Such a performance is made possible with unobstructed circular aperture telescopes, making these coronagraphs compelling options for probe-scale missions with 1.5\,m off-axis telescopes \citep{2010SPIE.7731E..67T,2010SPIE.7731E..68G,2013EPJWC..4715004B}.

Larger telescopes \citep[e.g. WFIRST-AFTA, ATLAST or ELTs][]{2013arXiv1305.5422S,2012OptEn..51a1007P,2013aoel.confE...8K} will provide smaller angular resolution and higher sensitivity for direct imaging and spectroscopy of exoplanets around a significant number of stars. However, their aperture present a complex geometry (secondary mirror central obstruction, primary mirror segmentation, and/or spiders) that produces diffraction features in the star point spread function (PSF), reducing the contrast performance for most coronagraphs and making broadband operation challenging. Several classical concepts \citep[e.g.][]{2000PASP..112.1479R,2002A&A...389..334A,2003ApJ...582.1147K,2005ApJ...633.1191M,2005ApJ...622..744G} have recently been revisited to explore new broadband solutions with high-contrast performance ($< 10^{-7}$) for arbitrary apertures (mainly large central obscuration).

Binary shaped pupils \citep{2003ApJ...582.1147K,2003ApJ...599..686V} reaches broadband contrasts lower than $10^{-7}$ with arbitrary pupils thanks to two-dimension pupil mask optimizations \citep{2011OExpr..1926796C}. The Four-Quadrant Phase Mask (FQPM) coronagraph \citep{2000PASP..112.1479R} has been improved in the context of unfriendly pupils with the addition of an optimized entrance pupil binary apodizer \citep{2013A&A...551A..10C}, leading to monochromatic contrasts better than $10^{-8}$. The Vector Vortex Coronagraph \citep[VVC, ][]{2005ApJ...633.1191M} has seen its theoretical perfect starlight extinction performance restored, thanks to the addition of a ring-apodized mask \citep{2013ApJS..209....7M} or an optimized binary shaped apodizer \citep{2013A&A...551A..10C} in the entrance pupil of the original scheme. For both VVC and FQPM concepts, the mitigation of the central obscuration effects and the efficiency increase in broadband light is also achieved by replicating these sectorized phase mask coronagraphs \citep{2000PASP..112.1479R,2005ApJ...633.1191M} in series \citep{2011A&A...530A..43G,2011OptL...36.1506M}.

Solving arbitrary aperture issues in coronagraphy by adapting Lyot stop geometry has been suggested for several concepts, such as classical Lyot-style concepts \citep{1933JRASC..27..265L,2001ApJ...552..397S} in the context of ELTs \citep{2005ApJ...626L..65S} or band-limited Lyot systems \citep{2002ApJ...570..900K,2007SPIE.6693E..46M,2008ApOpt..47..116B} in light of the WFIRST-AFTA coronagraph studies \citep{2013SPIE.8864E..12T}. The Phase-Induced Amplitude Apodization coronagraph and its derivatives \citep{2005ApJ...622..744G,2010ApJS..190..220G} also achieve contrast better than $10^{-7}$ over 2\% bandwidth with arbitrary shaped telescopes by introducing a Lyot stop with the same geometry as the pupil telescope. In addition to these monochromatic designs, further broadband optimizations are expected to improve the device performance \citep{2014ApJ...780..171G}. 

The Apodized Pupil Lyot Coronagraph \citep[APLC,][]{2002A&A...389..334A, 2003A&A...397.1161S} is the coronagraphic baseline adopted on most ground-based observatories like P1640, GPI, and SPHERE or for future ELT concepts to detect exoplanets at wide angular separations, thanks to its numerous properties: this system provides a $10^{-7}$ contrast, a moderate inner working angle (IWA $\sim 4\,\lambda_0/D$) and a high coronagraphic throughput for the off-axis target ($\simeq 50\%$) in the presence of telescopes with arbitrarily shaped apertures \citep{2005ApJ...618L.161S,2011ApJ...729..144S} and in particular, of ELTs \citep{2007A&A...474..671M,2009ApJ...695..695S}. The current generation of APLCs were designed using prolate apodizers, which are eigenfunctions associated with Lyot-style coronagraphic propagation for a given telescope aperture geometry. Using as optimization metric the contrast in the final focal plane inside the adaptive optics (AO) controllable region (i.e. AO dark zone), optimal apodizers for these APLCs were selected from the existing continuous set of prolate functions corresponding to the telescope geometry, and for a particular wavelength inside the bandpass. In addition, these APLCs also incorporated a simultaneous numerical optimization of the Lyot stop geometry and chromatic properties over the bandpass to deliver quasi-achromatic solutions over large bandpasses \citep{2011ApJ...729..144S}. These designs produce coronagraphic images that can advantageously be addressed with spectral deconvolution techniques \citep{2000PASP..112...91M,2002ApJ...578..543S,2008A&A...489.1345V} and advanced post-processing methods \citep{2007ApJ...660..770L,2012ApJS..199....6P,2012ApJ...755L..28S,2012MNRAS.427..948A} to optimally disentangle companions from speckle artifacts and retrieve their spectrum.

Althougth the final contrast was used as the optimization metric in the current generation of APLCs, their performance remains limited by the available degrees of freedom for the optimization, which is limited by the choice of prolate apodization and Lyot stop geometry. \citet{2013ApJ...769..102P} developed a different approach to introduce a fully numerical optimization of the apodization function for a given contrast target. This new approach relies on the APLC linear properties between the entrance pupil function and the coronagraphic electric field to find designs with optimal grey transmission apodization that generates coronagraphic images with high-contrast regions. 

In this case the apodizer is no longer selected from the continuous set of prolate eigenfunctions, but optimized directly for the set of constraints (contrast, IWA, etc.). This  approach is similar to shaped-pupil type optimizations \citep{2003ApJ...582.1147K,2003ApJ...599..686V,2011OExpr..1926796C} but in the APLC framework. This type of approach typically generates coronagraphic images with high-contrast regions in broadband light (dark zones), which were not observed with classical APLCs using prolate apodizations. This type of approach is well adapted to the presence of deformable mirrors (DMs) envisioned in the baseline of future exoplanet direct imagers \citep{2006ApOpt..45.5143S,2011SPIE.8151E..12K}, for wavefront control in the field of view within a region bounded by the maximum spatial frequency controllable by the DM. With this approach, \citet{2013ApJ...769..102P} found APLC solutions reducing starlight intensity down to a $10^{-10}$ level within a given area in the field of view in monochromatic light. 

In this paper we generalize these monochromatic solutions assuming gray transmission apodizer for broadband operations to enable spectroscopy of the detected companions. The large pupil obscuration in an on-axis telescope constitutes the main hurdle to overcome for broadband observations. Features such as support structures of the secondary mirror or segmentation of the primary mirror produce diffraction effects that can be mitigated in the Lyot plane \citep{2005ApJ...633..528S,2005ApJ...626L..65S}, allowing a coronagraph designed for rotationnally, axi-symmetric pupils to be used with unfriendly apertures \citep{2011ApJ...729..144S}. Such a strategy has been adopted for GPI and P1640 \citep{2008SPIE.7015E..31M,2011PASP..123...74H}. Pupil remapping techniques have also been proposed with the phase-induced amplitude apodization \citep{2005ApJ...622..744G} to mitigate central obscuration and spider vanes. This method was adopted on SCExAO with large support structures \citep{2009SPIE.7440E..0OM}. The Active Correction of Aperture Discontinuities (ACAD) method is another, recent and promising pupil remapping technique that is applicable to any type of coronagraph to mitigate diffraction struts with the use of two sequential deformable mirrors \citep{2013ApJ...769..102P}. We here develop solutions for circularly symmetric obstructed pupils to overcome the diffraction effects of central obscuration in broadband high-contrast imaging. 

In Section \ref{sec:optimization}, we recall the linear properties of APLCs and develop broadband designs for a given set of constraints (contrast, IWA, etc.) by defining a linear optimization problem. In Section \ref{sec:results}, we give a first application with some solutions for GPI and analyze their performance (contrast, IWA, throughput, and achromaticity). In Section \ref{sec:sensitivity}, we analyze the solutions in terms of sensitivity to low-order aberration and underline their attractiveness since they can be rendered quasi-insensitive to effects of star jitter or stellar diameter. We also illustrate our method by presenting designs that could potentially be used for upgrades of existing APLC facilities or for future high-contrast instruments.

\section{Optimization problem}\label{sec:optimization}
\subsection{Coronagraphic operator $\mathcal{C}_\lambda$ and linearity properties}
\begin{figure}[!ht]
\centering
\resizebox{\hsize}{!}{
\includegraphics{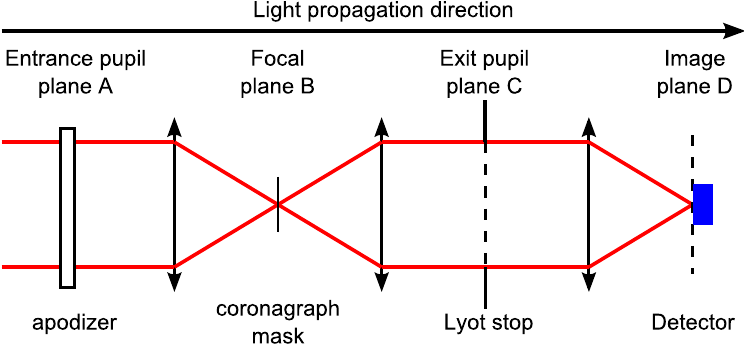}
}
\caption{Optical layout of the Apodized Pupil Lyot Coronagraph. This device involves four different plans in which are inserted the coronagraphic components with the entrance pupil apodizer in plane A, the opaque mask in the intermediate focal plane B, the Lyot stop in the relayed pupil plane C and finally, the camera onto which the coronagraphic image is formed.}
\label{fig:APLC layout}
\end{figure}

The APLC is a Lyot-style coronagraph which involves four planes (A, B, C, D) and combines an entrance pupil apodization in plane A, a downstream FPM in plane B, and a Lyot stop in the relayed pupil plane C, to form the coronagraphic image of an unresolved, on-axis bright star on a detector located in the re-imaged focal plane D, see Figure \ref{fig:APLC layout}. We recall the formalism of the APLC and establish the expression of an operator $\mathcal{C}_{\lambda}$ that linearly relates the entrance pupil function $P$ and the coronagraphic field at a given wavelength $\lambda$ within the bandpass $\Delta\lambda$ centered at central wavelength $\lambda_0$. 

In the following, we consider axi-symmetric pupils and masks, allowing us to use radial functions to describe the formalism, with no loss of generality, see e.g. \citet{2011OExpr..1926796C}. The optical layout of the coronagraph is such that the complex amplitudes of the electric field in two successive planes are related by a Hankel transform. We call $\widehat f$ the Hankel transform of a function $f$:
\begin{equation}
\widehat{f}(\xi)=\int_{0}^{+\infty} f(r)\,J_0(2\pi\,r\xi)\,2\pi\,r\,dr\,,
\label{eq:Hankel}
\end{equation}
where $J_0$ denotes the zero-order Bessel function of the first kind and $\xi$ represents the focal plane coordinate expressed in units of angular resolution ($\lambda_0/D$).

A telescope aperture with central obstruction of respective diameters $D$ and $d$ forms the coronagraph entrance pupil of the system. Assuming the presence of an apodizer in the same plane, the entrance pupil function $P$ is defined as non null for $d/2 < r < D/2$. The electric field $\Psi_A$ in the entrance pupil plane A of the system is simply given by
\begin{equation}
\Psi_A(r, \lambda)=P(r)\,.
\label{eq:Psi_A}
\end{equation} 

A FPM of angular diameter $m$ is introduced in the following focal plane B. Its transmission is given by $1-M$ with $M(\xi) = 1$ for $\xi < m/2$ and 0 otherwise. At the operating wavelength $\lambda$, the electric field $\Psi_B$ in the following focal plane is given by
\begin{equation}
\Psi_B(\xi, \lambda)=\frac{\lambda_0}{\lambda} \widehat{\Psi}_A\left ( \frac{\lambda_0}{\lambda}\xi, \lambda \right ) [1-M(\xi)]\,.
\label{eq:Psi_B0}
\end{equation}
By developing $\widehat{\Psi}_A(\xi\lambda_0/\lambda, \lambda)$ in the second term, we obtain 
\begin{equation}
\begin{split}
\Psi_B(\xi, \lambda) &=\frac{\lambda_0}{\lambda}\left [\widehat{P}\left (\frac{\lambda_0}{\lambda}\xi \right )\right .\\
&\quad\left . -M(\xi)\int_{0}^{+\infty} P(u)\,J_0\left (2\pi\,u\frac{\lambda_0}{\lambda}\xi\right )\,2\pi\,u\,du\,\right ].
\end{split}
\label{eq:Psi_B}
\end{equation}

The complex amplitude $\Psi_C$ in the relayed pupil plane C is obtained by Hankel transform of the previous expression followed by Lyot stop filtering:
\begin{equation}
\Psi_C(r, \lambda)=\frac{\lambda_0}{\lambda}\widehat{\Psi}_B\left ( \frac{\lambda_0}{\lambda}r, \lambda \right ) L(r)\,,
\label{eq:Psi_C0}
\end{equation}
where $L$ is the Lyot stop function defined as $L(r)=1$ for $d_S/2 < r < D_S/2$ and 0 otherwise, where $d_S$ and $D_S$ denote the sizes of the inner and outer edge of the aperture stop, with $d_S \geq d$ and $D_S \leq D$. The complex amplitude $\Psi_C$ can be rewritten as
\begin{equation}
\begin{split}
\Psi_C(r, \lambda)=&\left [P(r)\right .\\
&\left .- \left (\frac{\lambda_0}{\lambda}\right )^2\int_0^{+\infty} P(u)\,K_M\left (\frac{\lambda_0}{\lambda}u, \frac{\lambda_0}{\lambda}r \right )\,u\,du \right ] L(r)\,,
\end{split}
\label{eq:Psi_C}
\end{equation}
where the second term describes the wave diffracted by the FPM. This term is then expressed using the Kernel convolution $K_M$ with
\begin{equation}   
K_M(u, r)=(2\pi)^2\int_0^{m/2} J_0\left (2\pi\,\eta \,u\right )\,J_0\left (2\pi\,\eta\,r\right )\,\eta\,d\eta\,.
\label{eq:kernel}
\end{equation}

The coronagraphic electric field $\Psi_D$ is finally obtained with a scaled Hankel transform of the Lyot plane field:
\begin{equation}
\Psi_D(\xi, \lambda)=\frac{\lambda_0}{\lambda} \widehat{\Psi}_C\left ( \frac{\lambda_0}{\lambda}\xi, \lambda \right )\,.\\
\label{eq:Psi_D0}
\end{equation}
Developing this equation makes appear a relation between the electric field $\Psi_D$ and the entrance pupil function $P$ that can be underlined by means of an operator $\mathcal{C}_\lambda$ as follows
\begin{equation}
\begin{split}
\mathcal{C}_\lambda[P(r)](\xi) &=\Psi_D(\xi, \lambda)\\
&=\left (\frac{\lambda_0}{\lambda} \right )\left [\int_{d_S/2}^{D_2/2}P(r)\,J_0\left (2\pi\,r\frac{\lambda_0}{\lambda}\xi\right )\,2\pi\,r\,dr\right .\\
 &\quad\left .-\left (\frac{\lambda_0}{\lambda}\right )^2\int_{0}^{m/2} \widehat{P}\left (\frac{\lambda_0}{\lambda}\eta \right )\,K_S\left (\frac{\lambda_0}{\lambda}\xi, \frac{\lambda_0}{\lambda}\eta\right )\,\eta\,d\eta\,\right ],
\end{split}
\label{eq:Psi_D}
\end{equation}
where the behavior of the Lyot stop on the wave diffracted by the FPM is described by the second term, expressed using the Kernel convolution $K_S$:
\begin{equation} 
K_S(\xi, \eta)=(2\pi)^2\int_{d_S/2}^{D_S/2} J_0\left (2\pi\,r\,\xi\right )\,J_0\left (2\pi\,r\,\eta\right )\,r\,dr\,.
\end{equation} 
The Kernel convolutions $K_M$ and $K_S$ reflect the filtering effect induced by the FPM and the Lyot stop. An analytical closed form for these Kernels is given in \citet{2003A&A...397.1161S}. 

\citet{2013ApJ...769..102P} have previously derived an expression for $\mathcal{C}_{\lambda}$ (see Eq. (2) of their paper\footnote{Note the typo in \citet{2013ApJ...769..102P}'s Eq. (3): Kernel convolution captures the effect of the Lyot stop instead of the FPM.}) at $\lambda_0$ and in the absence of Lyot stop undersizing ($D_S=D$ and $d_S=d$). Eq. (\ref{eq:Psi_D}) constitutes an extension of their expression since our operator $\mathcal{C}_{\lambda}$ enables Lyot stop undersizing and expresses the coronagraphic electric field at any wavelength $\lambda$ within the spectral bandpass $\Delta\lambda$ centered at $\lambda_0$.

The coronagraphic intensity over the bandpass $\Delta\lambda$ can be deduced from Eq. (\ref{eq:Psi_D}) by means of a second operator $\mathcal{I}_{\Delta\lambda}$ as
\begin{equation}
\mathcal{I}_{\Delta\lambda}[P(r)](\xi)=\frac{1}{\Delta\lambda}\int_{\Lambda} \left [\mathcal{C}_{\lambda}[P(r)](\xi)\right ]^2\,d\lambda\,,
\label{eq:I_Dbroad}
\end{equation}
where $\Lambda$ defines a set of wavelengths $\lambda$ such that $|\lambda-\lambda_0|< \Delta\lambda/2$.

\citet{2013ApJ...769..102P} developed APLC solutions to generate coronagraphic star image with dark hole $\mathcal{D}$ in monochromatic light observations, relying on the linear relation between the coronagraphic electric field and the apodization at $\lambda_0$ in Eq. (\ref{eq:Psi_D}). Optimized at a single wavelength, these coronagraphic designs do not form a star image with broadband dark region in white light observations.

Our goal is to optimize the entrance pupil apodization for an APLC to reach a given contrast $C$ over a specific area  $\mathcal{D}$ in a broadband coronagraphic image. Eq. (\ref{eq:I_Dbroad}) is a quadratic form between the coronagraphic broadband intensity $\mathcal{I}_{\Delta\lambda}$ and the entrance pupil apodizer $P$, which makes the search of the optimal apodization difficult. To circumvent this problem and develop novel APLC broadband solutions, we propose a multi-wavelength approach based on Eq. (\ref{eq:Psi_D}) which relates the entrance pupil apodization and the coronagraphic electric field at a given wavelength. 

\subsection{Numerical optimization of the apodizer}
The parameters for the optimization of a classical APLC for a given telescope geometry are: mask size, bandpass, Lyot stop geometry, wavelength at which the eigenfunction is defined, and IWA. Quantities such as the contrast and throughput can be derived independently, and the overall APLC design can be selected to reach a certain level of contrast and throughput. However these important quantities are not part of the optimization problem.

Here our approach is to define a numerical optimization problem, including all the classical APLC parameters, plus the throughput, contrast, and outer working angle (OWA). The goal is to find the apodizer transmission profile for the APLC that reaches a contrast target $C$ in the broadband coronagraphic PSF within the search area $\mathcal{D}$, as a function of other parameters (Lyot stop geometry, bandpass, mask size). Among the solutions that can meet these criteria, the most desirable solution is the one that offers the best transmission to provide the highest off-axis throughput.

Since the operator $\mathcal{C}_{\lambda}$ is linear at a given $\lambda$, the optimal pupil apodization for the APLC can be found by solving the following linear optimization problem:
\begin{subequations}
      \renewcommand{\theequation}{\theparentequation\alph{equation}}
\begin{alignat}{2}
& \max_{\{pk\}}[P(r)] \quad \text{for} \quad d/2<r<D/2 \quad \label{subeq:maxminP} \text{under the contraints:}\\
& \left |\mathcal{C}_{\lambda}[P(r)](\xi) \right | < \sqrt{10^{-C}} \quad \text{for} \quad
\begin{dcases}
\rho_i < \xi < \rho_o\\
|\lambda-\lambda_0| < \Delta\lambda/2 \label{subeq:C} 
\end{dcases}\\
& \max_r[P(r)] =1 \label{subeq:maxP}\\
& \left |\frac{d}{dr}[P(r)] \right | < b. \label{subeq:maxdP/dr}
\end{alignat}
\end{subequations}
We explain the meaning of theses equations in the following.

\noindent (\ref{subeq:maxminP}) The apodizer throughput is a quadratic function of the field amplitude and maximizing it requires the solution of a nonlinear optimization problem (as described in \citet{2003ApJ...590..593V} and recalled in \citet{2013ApJ...769..102P}). In the context of our linear optimization problem, we use as an approximation a linear constraint of the amplitude at each pupil point as a criterion in an attempt to maximize the apodization throughput.\\
(\ref{subeq:C}) The absolute value of the coronagraphic electric field is constrained at all the wavelengths over the bandpass to achieve a contrast C over the area $\mathcal{D}$ in broadband light. This region is defined between its inner and outer edges, repsectively $\rho_i$ and $\rho_o$.\\ 
(\ref{subeq:maxP}) The maximum value of the apodization function is set to one to exclude the null function in the resolution of our problem.\\
(\ref{subeq:maxdP/dr}) We introduce a constraint on the absolute value of the derivative of $P$ with respect to the radial coordinate to limit the oscillations within the function $P$ and avoid bang-bang solutions that make the apodizer manufacturing very challenging. The constant $b$ is defined as the derivative limit.

We solve this problem by using a linear programming solver \citep{Vdb09} and in the following, we rely on the routine provided with the Mathematica software \citep{Mathematica9.0}. 

The entrance pupil apodization function can be represented in a number of ways, including directly sampled as independent field amplitude pixel values, or decomposed on a modal basis. Here we use a weighted sum of radial Bessel modes as a pupil representation, and we adopt an expression for the entrance pupil function in the form:
\begin{equation}
P(r)=\sum_{k=0}^{N_{modes}} p_k\,J_0\left (\frac{r}{\alpha_k}\right )\,,
\end{equation}
where $\alpha_k$ represents the $k$th zero of the zero-order Bessel function of first kind $J_0$. Bessel functions of higher order could be used to decompose $P$ as proposed by \citet{2013ApJ...769..102P}; we realized tests with different orders and found similar results for the first few orders. With this modal representation, the numerical optimization seeks the coefficients $p_k$, instead of the pixel values themselves, which helps to reduce the dimensionality of the problem and accelerates the computation. As a sanity check, we have verified that both pixel-based or modal-based representations of the apodizers both lead to very similar results.

Multi-spectral constraint on the coronagraphic electric field is used in the optimization problem to find solutions producing broadband PSF dark zones with the specified contrast. The problem is solved here by using a number of five wavelengths for the spectral constraints in 10\% and 20\% bandpasses. We performed some tests with different number of wavelengths that are equally spatially sampled one to another within the bandpass for our optimization. Similar results are found for five wavelengths and beyond. We have also computed the coronagraphic electric field at wavelengths that are unconstrained in the optimization procedure and we have verified that the coronagraph contrast at the wavelengths within the specified bandpass meets our problem constraints.

\subsection{Comparison with Prolate apodizations}
Before investigating novel designs, we first ran some tests to show the ability of our method to recover the APLC designs using the classical prolate apodization as defined by \citep{2011ApJ...729..144S}. Our broadband optimization is made under the following assumptions.

We consider a circular aperture with 20\% central obscuration (expressed in pupil radius ratio) and a FPM of radius $m/2=3.31\,\lambda_0/D$, corresponding to the relative dimensions of our reflective mask installed in the High Contrast Imager for Complex Aperture Telescope (HiCAT) testbed \citep{2013SPIE.8864E..1KN,2014arXiv1407.0980N} 
and inherited from the Lyot project \citep{2004SPIE.5490..433O}. The standard method to generate APLC prolate apodizations involves a Lyot stop with the same shape as the entrance pupil \citep{2002A&A...389..334A, 2003A&A...397.1161S} and therefore, no oversizing of the Lyot stop obstruction is introduced in our optimization. The prolate apodization for the mask size and the aperture geometry mentioned above gives a $\sim 10^{-8}$ contrast at 5.5\,$\lambda_0/D$. We set a contrast target $C=8$ over a 20\% bandpass within the search area $\mathcal{D}$ defined by $\rho_i=5.5\,\lambda_0/D$ and $\rho_o=48\,\lambda_0/D$ to look for the prolate apodization solution. A large number of modes $N_{modes}=48$ is chosen to control the spatial frequencies in the electric field up to the outer edge of $\mathcal{D}$, since the generation of the prolate apodization does not account on any specific search area delimitations. Finally, since the prolate apodization does not present large amplitude fluctuations, we set the constraint of the derivative limit to a small value $b=5$.

\begin{figure*}[!ht]
\centering
\resizebox{\hsize}{!}{
\includegraphics[height=5.0cm]{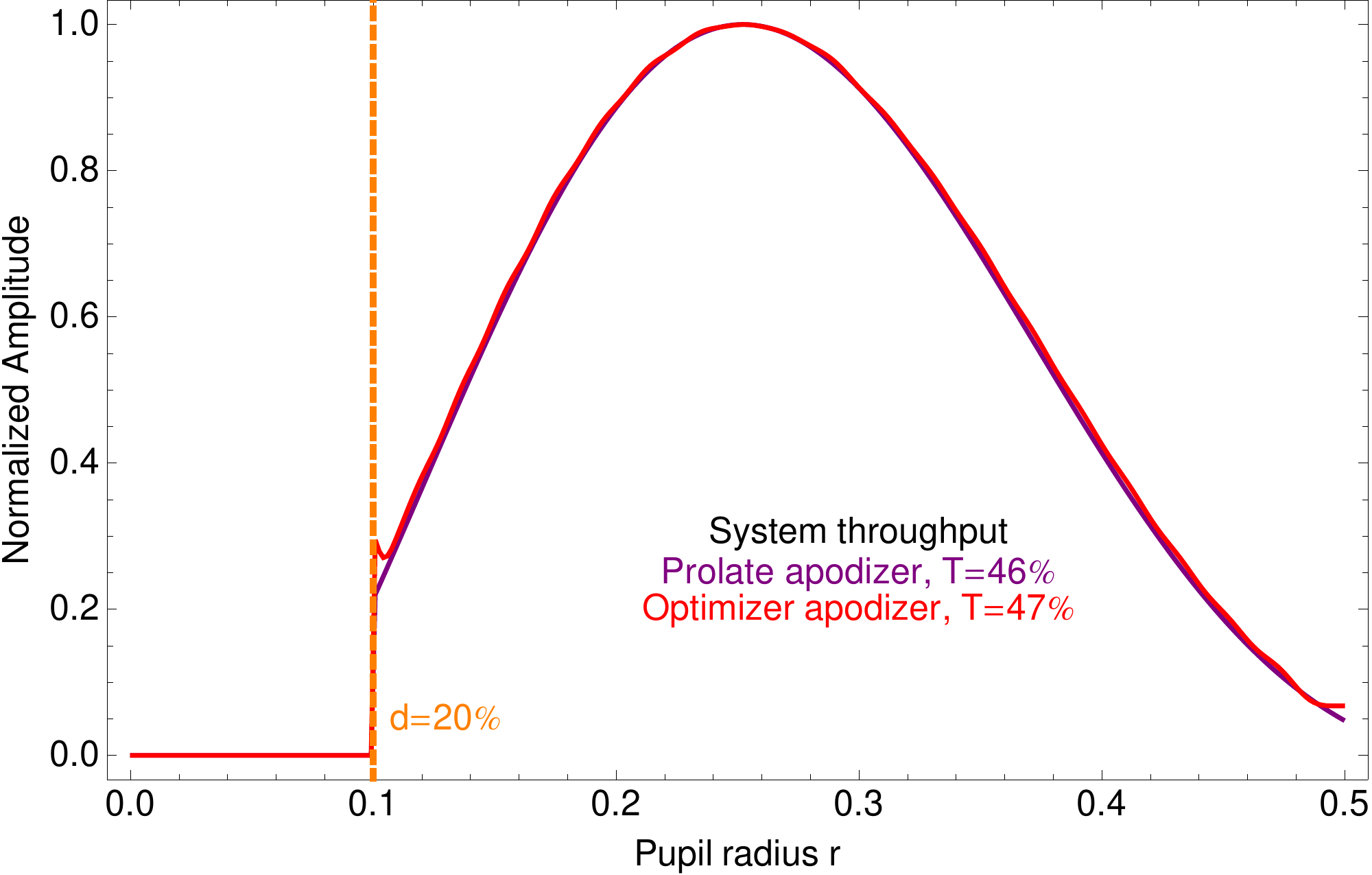}
\includegraphics[height=5.0cm]{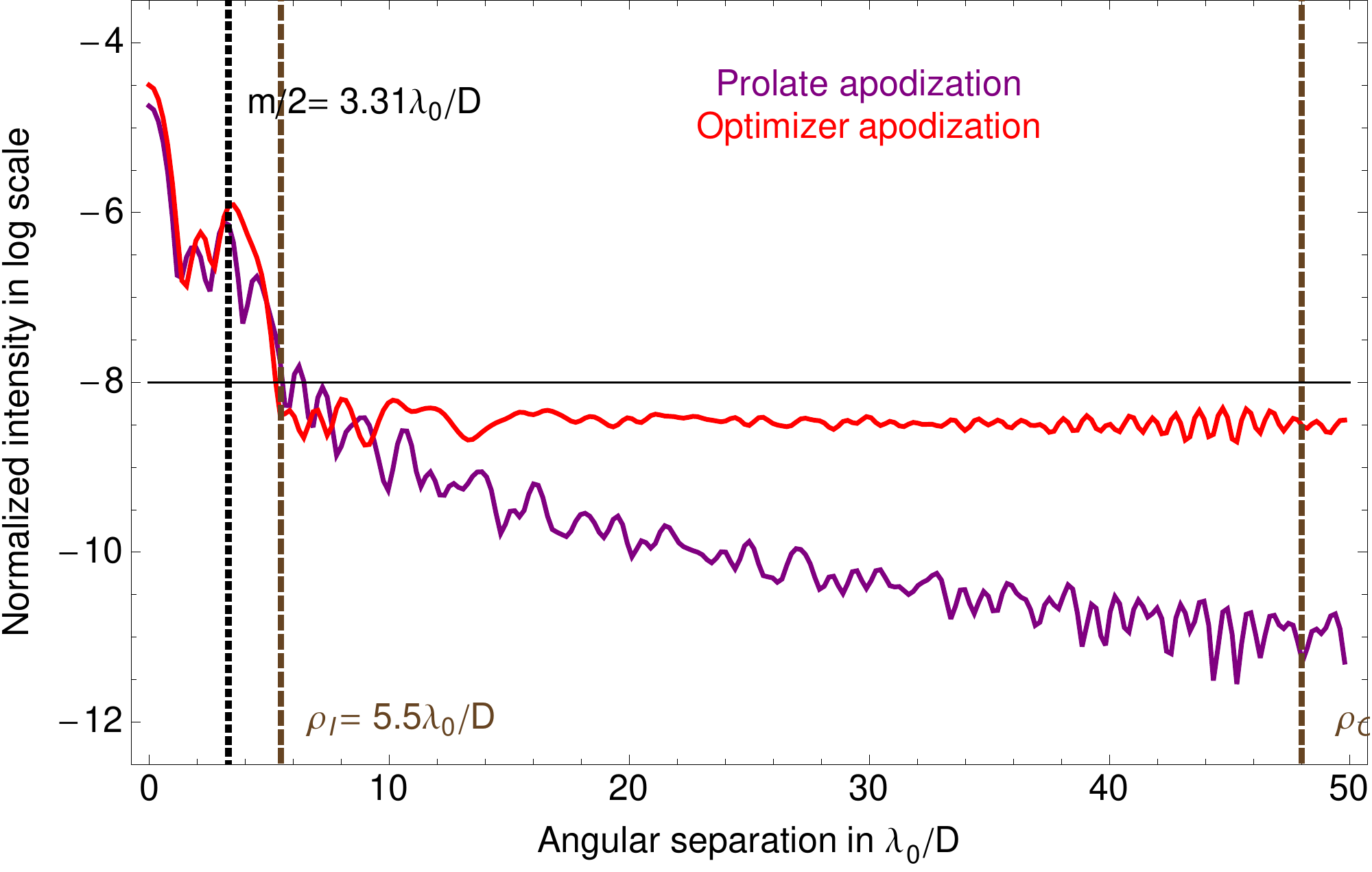}
}
\caption{Comparison of APLC designs with apodization using a prolate function and our linear optimization approach. \textbf{Left}: radial amplitude profiles of the prolate apodization calculated at $\lambda_1=1.05\,\lambda_0$ following the method in \citet{2011ApJ...729..144S} (purple) and the apodization found with the linear programming optimizer (red) for a 20\% bandpass, a pupil with 20\% central obscuration (in radius ratio) and a FPM ($m/2=3.31\,\lambda_0/D$). The orange line delimits the radius of the pupil obscuration. \textbf{Right}: Coronagraphic intensity profiles reached by APLC with the apodizers shown on the left plot. The brown dashed lines represent the bounds of the search area $\mathcal{D}$ ($\rho_i=5.5\,\lambda_0/D$ and $\rho_o=48\,\lambda_0/D$) and the black dot line delimits the FPM radius. By properly defining our problem, our linear optimization method finds an APLC solution with apodizer close to the optimal prolate function found by \citet{2011ApJ...729..144S}, showing the ability of our approach to recover existing designs.}
\label{fig:ComparisonProlate}
\end{figure*}

Figure \ref{fig:ComparisonProlate} shows the apodization and the resulting coronagraphic intensity profile for the linear programming solution and the prolate apodization calculated at the wavelength $\lambda_1=1.05\,\lambda_0$ for optimal broadband extinction, following \citet{2011ApJ...729..144S}. This first example confirms the ability of our linear optimization method to achieve the prolate apodization solution previously found by \citet{2011ApJ...729..144S}. 
Also, it is important to note that for such linear programming optimizer, the existence of a solution is not guaranteed for a given set of parameters, as opposed to the case of a classical APLC where a prolate apodizer (eigenfunction) always exists for a given geometry and combination of mask size and wavelength.

\section{Application to the Gemini Planet Imager}\label{sec:results}
As an illustration of our approach, we present an application of the APLC design optimizations by considering the parameters of GPI to investigate future possible ugrade of the instrument.

\subsection{Parameters}
We assume a circular entrance pupil with a 14\% central obstruction ratio and we select the FPM of relative radius $m/2=3.48\,\lambda_0/D$ in H-band. For the coronagraph design, we set our contrast target to $10^8$ in the dark zone over a 20\% bandpass, corresponding to one order of magnitude higher contrast than the specification for the GPI coronagraphs \citep{2008SPIE.7015E..31M,2009SPIE.7440E..0RS,2011ApJ...729..144S}. The dark-zone region has an outer bound $\rho_o=20\,\lambda_0/D$, corresponding to the maximum spatial frequency that can be controlled with the GPI deformable mirror.

In our simulations, we set a pupil size of $N=300$ pixels for camera images with maximum spatial frequency of 50 element resolutions (or cycles/pupil). From this last number, we set $N_{modes}=48$. In the following, we explore the entire parameter space by considering the Lyot stop inner diameter (ID) oversizing factor and the inner radius $\rho_i$ of the search area $\mathcal{D}$ by solving the optimization problem described above. The ID oversizing factor is defined as the ratio between the central obscuration diameters of the Lyot stop and the aperture $d_s/d$. A similar undersizing factor can also be defined as the ratio of the outer diameters of the Lyot stop and the aperture $D_S/D$. We ran some tests and did not find any optimum for this parameter, which is in accordance with the results described in \citet{2011ApJ...729..144S}. We set the constraint on the derivative $b$ to 5 to avoid bang-bang solutions that typically have very low throughput, and to avoid the hard-to-manufacture intermediate situation with a continuous transmission profile but very high oscillatory behavior that is obtained if larger derivative values are tolerated. 

\subsection{Parameter space exploration}
\begin{figure*}[!ht]
\centering
\resizebox{\hsize}{!}{
\includegraphics{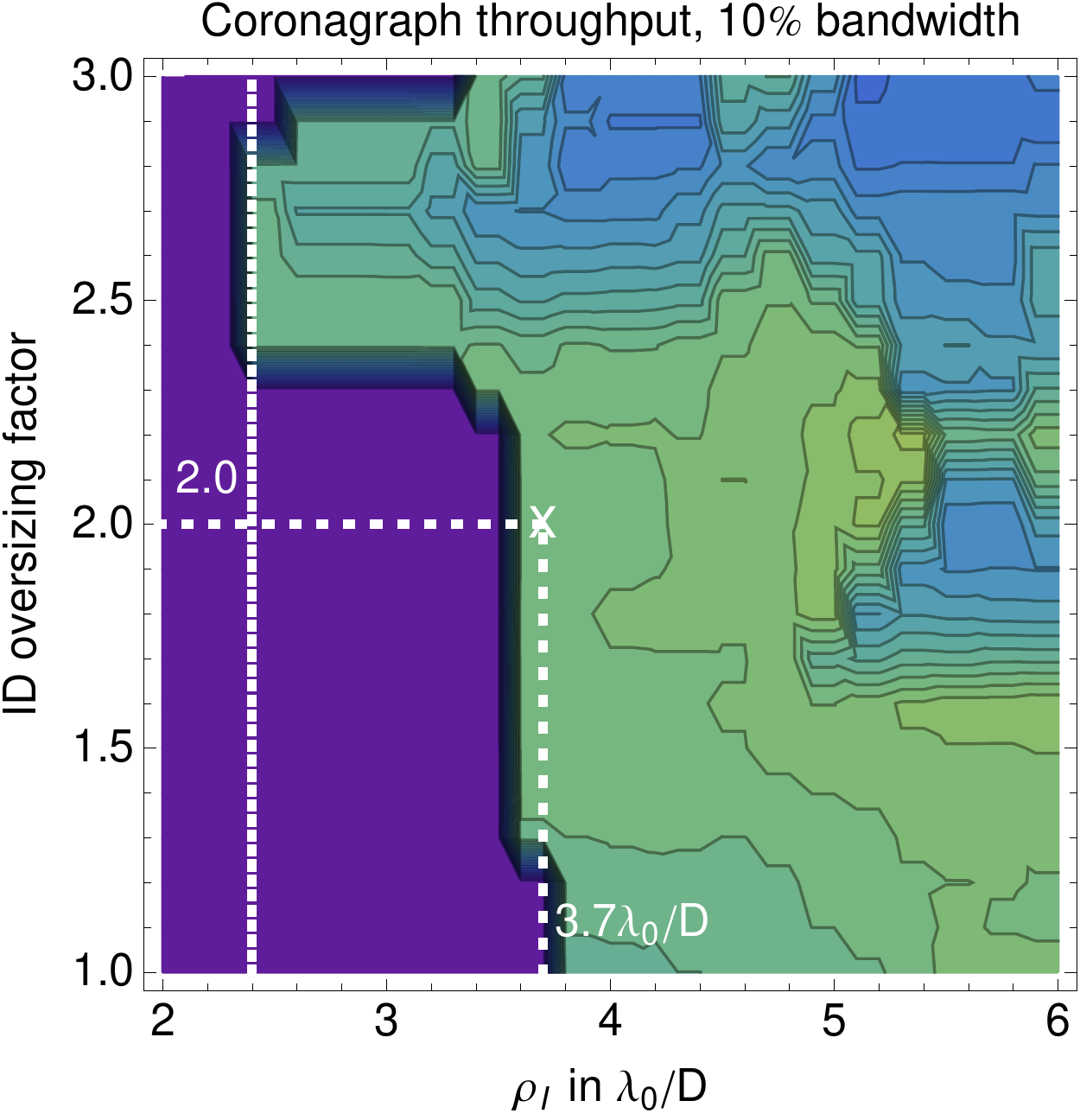}\hspace{5cm}
\includegraphics{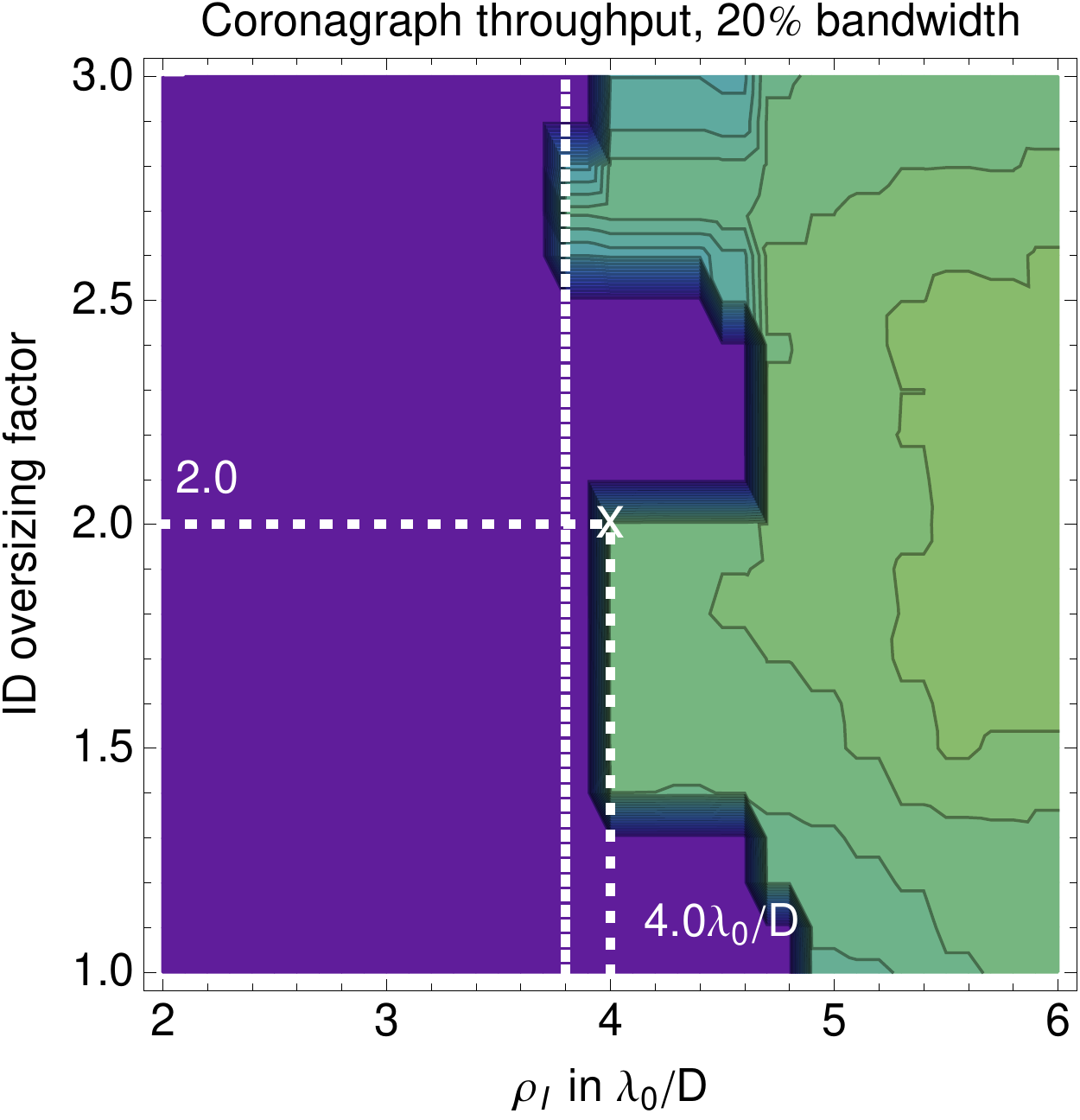}\hspace{2cm}
\includegraphics{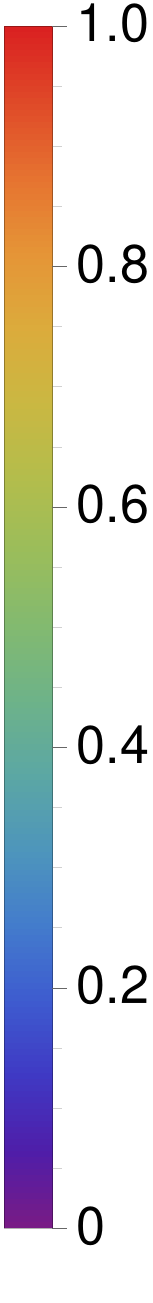}
}
\caption{Coronagraph throughput as a function of the search area inner radius $\rho_i$ and the Lyot stop oversizing factor for a mask radius $m/2 = 3.48\,\lambda_0/D$, a circular pupil with 14\% central obscuration in radius ratio, a search area outer radius $\rho_o = 20\,\lambda_0/D$, and a contrast of $C=8$. Coronagraph throughput is represented for a 10\% bandwidth (left panel) and 20\% bandwidth (right). Note that one point in these plots corresponds to an APLC design with a value associated to the coronagraph throughput criterion, and that regions with zero throughput (in purple) correspond to cases where the linear program does not have have a solution. Two different regimes appear in these contour plots, delimited by white dotted lines. No solution is found at any ID oversizing factor (regime 1) for inner edges smaller than 2.4\,$\lambda_0/D$ (left) and 3.8\,$\lambda_0/D$ (right). On the opposite (regime 2), solutions exist for any ID oversizing factor and islands in the coronagraph throughput can be distinguished in the contour plot, underlining the existence of optimal values of ID for a given $\rho_i$. The cross marks denote the selected solutions for Lyot stop oversizing factor $ID=2.0$ and showed in Fig. \ref{fig:GPI_solution}.}
\label{fig:rho_i_vs_ID}
\end{figure*}

Figure \ref{fig:rho_i_vs_ID} represents the throughput of the complete coronagraphic system, including the apodizer and Lyot stop, as a function of the search area inner radius $\rho_i$ and the Lyot stop ID oversizing factor for the GPI mask mentioned above. Two contour plots are displayed, representing the 10\% and 20\% bandwidths (left and right respectively). The parameters are listed in Table \ref{table:rho_i_vs_ID}.

\begin{table}[!ht]
\caption{Parameters for the panels showed in Figure \ref{fig:rho_i_vs_ID}.}
\centering
\begin{tabular}{c c c c}
\hline\hline
Aperture & Bandwidth & Mask radius & Contrast\\
obstruction (\%) & (\%) & $m/2$ ($\lambda_0/D$) & target C\\
\hline
14 & 10 and 20 & 3.48 & 8\\
\hline
\end{tabular}\\
\label{table:rho_i_vs_ID}
\end{table}

In each plot, two different regimes are observed, delimited by the existence or not of solutions to the optimization problem with our constraints. This limit, found at 3.8\,$\lambda_0/D$ for 20\% bandwidth, is pushed down to 2.4\,$\lambda_0/D$ for a 10\% bandwidth, underlining that enlargement of the solution space is possible for a given constraint on the apodizer derivative at the cost of a bandwidth decrease. This parameter space exemplifies the very important trade-off between bandwidth and IWA in the optimization of the apodizer.

Relaxing the constraint on the apodizer derivative (larger $b$) can be used to find solutions for a wide spectral bandpass ($> 10\%$) and a small search area inner bound ($< 4\,\lambda_0/D$) but in return, the optimizer apodizer will present large and fast amplitude variations over the pupil; these apodizer configurations have very low throughput, and the fast oscillations are most likely very hard to manufacture, e.g. with microdots technology \citep{2009A&A...495..363M,2009SPIE.7440E..40S}.

\subsection{GPI solution using an existing mask of radius $m/2=3.48\,\lambda_0/D$}
Assuming an $ID=2$ oversizing factor for the Lyot stop obstruction, we seek two solutions for $10\%$ and $20\%$ bandwidths with the smallest inner edge of the dark region and the highest coronagraphic throughput observed in Figure \ref{fig:rho_i_vs_ID} plots, see white cross marks. The parameters of these two solutions are recalled in Table \ref{table:GPI_solution}.

\begin{table}[!ht]
\caption{Parameters for the designs showed in Figure \ref{fig:GPI_solution}.}
\centering
\begin{tabular}{c c c c}
\hline\hline
Bandwidth & Dark region & Contrast & Coronagraph\\
(\%) & inner edge $\rho_i$ ($\lambda_0/D$) & target C & throughput (\%)\\
\hline
10 & 3.7 & 8 & 39\\
20 & 4.0 & 8 & 39\\
\hline
\end{tabular}\\
\label{table:GPI_solution}
\tablecomments{FPM of radius $m/2=3.48\,\lambda_0/D$ for an aperture with 14\% central obstruction and a Lyot stop with 28\% central obstruction in pupil radius ratio.}
\end{table}

\begin{figure*}[!ht]
\centering
\resizebox{\hsize}{!}{
\includegraphics{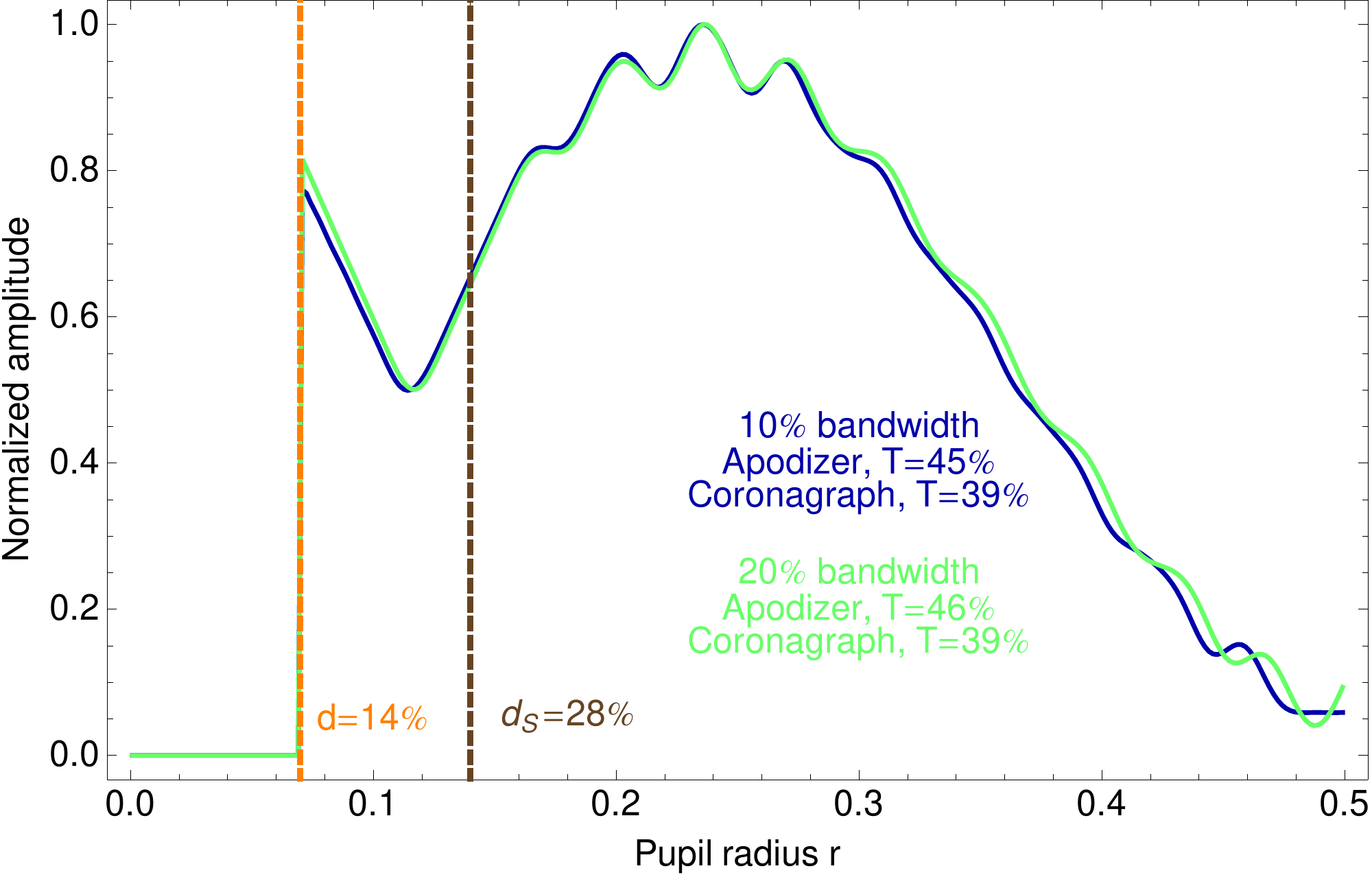}
\includegraphics{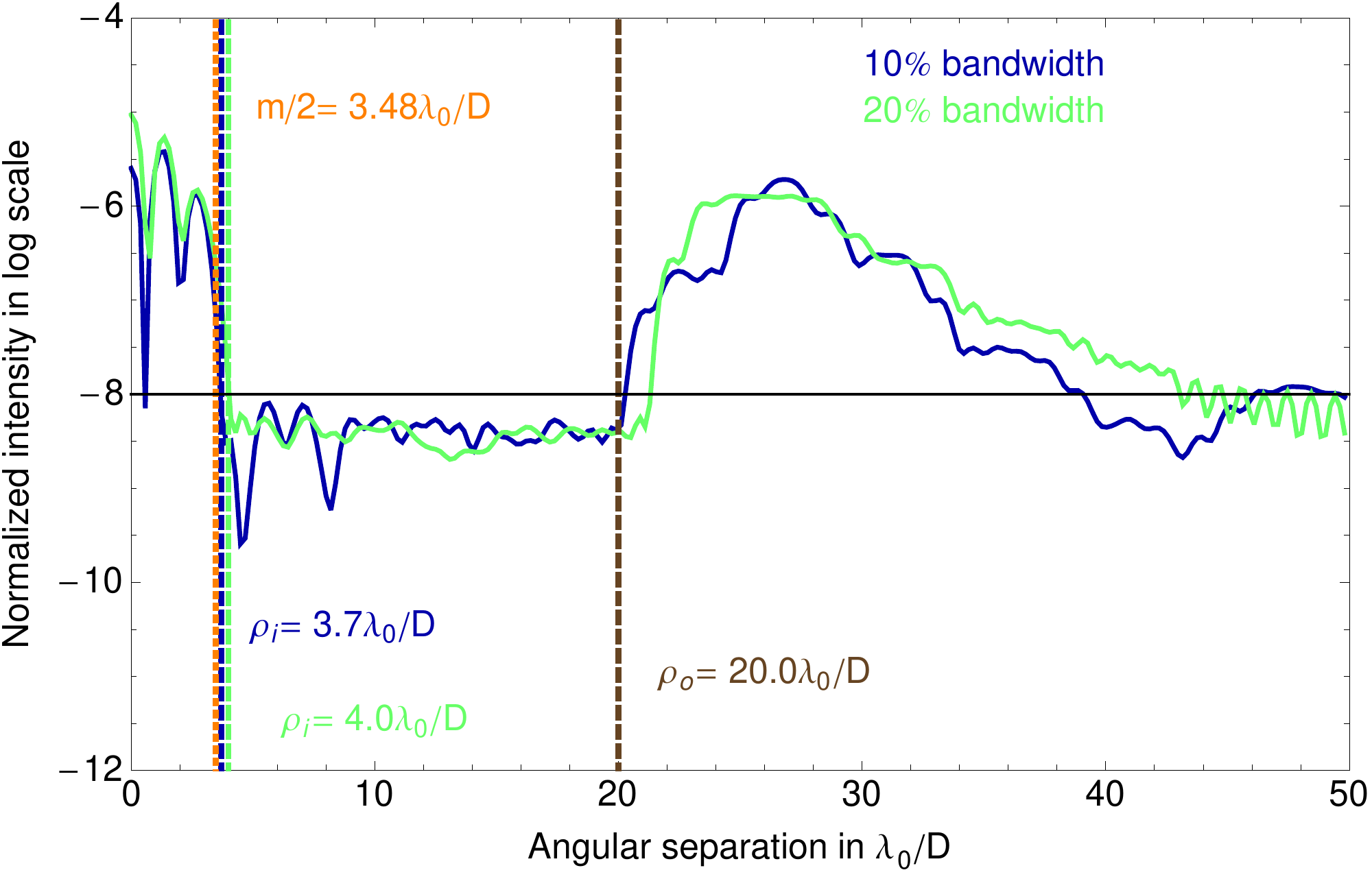}
}
\caption{Illustration of optimized new solutions for the GPI configuration and using an existing GPI FPM. \textbf{Left}: Radial amplitude profiles of the apodizer for 10\% bandwidth (blue) and 20\% bandwidth (green). These designs are optimized for an aperture with 14\% central obscuration in pupil radius ratio (orange vertical line), a FPM radius $m/2=3.48\,\lambda_0/D$, a Lyot stop ID oversizing factor of 2 (corresponding to a size $d_S=28\%$ in aperture radius ratio, brown vertical line), an inner bound search area $\rho_i=$3.7 and 4.0\,$\lambda_0/D$ respectively, and a common outer bound search area $\rho_o=20\,\lambda_0/D$. \textbf{Right}: Radial intensity profiles of the broadband coronagraphic images with these apodizers over the respective bandwidth. The focal mask radius, the inner bound of the search area for 10\% and 20\% bandwidth, the outer bound of the dark region are delimited with orange, blue, green, and brown dashed vertical lines. The average contrast over the spectral band in the dark region is below $10^{-8}$ (black horizontal line) for both designs.}
\label{fig:GPI_solution}
\end{figure*}

Since these solutions are based on re-using an existing FPM in GPI (in this case a K-band FPM considered for use at H-band), the mask size is not part of this optimization. We find two designs with 39\% coronagraph throughput for $\rho_i = 3.7$ and $4.0\,\lambda_0/D$ with 10\% and 20\% bandwidth. The apodizers and their coronagraphic profiles are represented in Figure \ref{fig:GPI_solution}. Our designs provide coronagraphic images with $10^8$ contrast dark regions in white light for two different bandwidths.

\subsection{Reduction of the inner edge of the dark region}
The $\rho_i = 2.4\,\lambda_0/D$ bound is the shortest inner edge of the dark hole obtained with our 10\% bandwidth solution for a mask radius $m/2 = 3.48\,\lambda_0/D$ and a Lyot stop with 28\% central obscuration in pupil radius ratio, based on the actual diameter of K-band mask and the current Lyot stop geometry of the GPI instrument. As part of our parameter space study we found that solutions exist with $\rho_i$ (i.e. the inner edge of the dark zone) smaller than the mask radius. In this configuration the actual IWA of the instrument (i.e. defining the ability to detect an object) is entirely determined by the geometry of the FPM, and not that of the coronagraph dark zone (note that the actual instrumental dark zone also depends on wavefront control and diffraction effects from secondary support structures and/or segment gaps, not discussed here). Producing dark holes with an inner bound smaller than the projected FPM radius is a fundamentally new property of APLCs with numerical optimization of the apodizer. We find that the existence of this new type of solution is also strongly dependent on the entrance pupil central obstruction. These solutions are very interesting in practice with the benefit of being quasi insensitive to tip tilt and more generally low-order aberrations. We thus investigate this aspect in more details below.

\begin{figure}[!ht]
\centering
\resizebox{\hsize}{!}{
\includegraphics{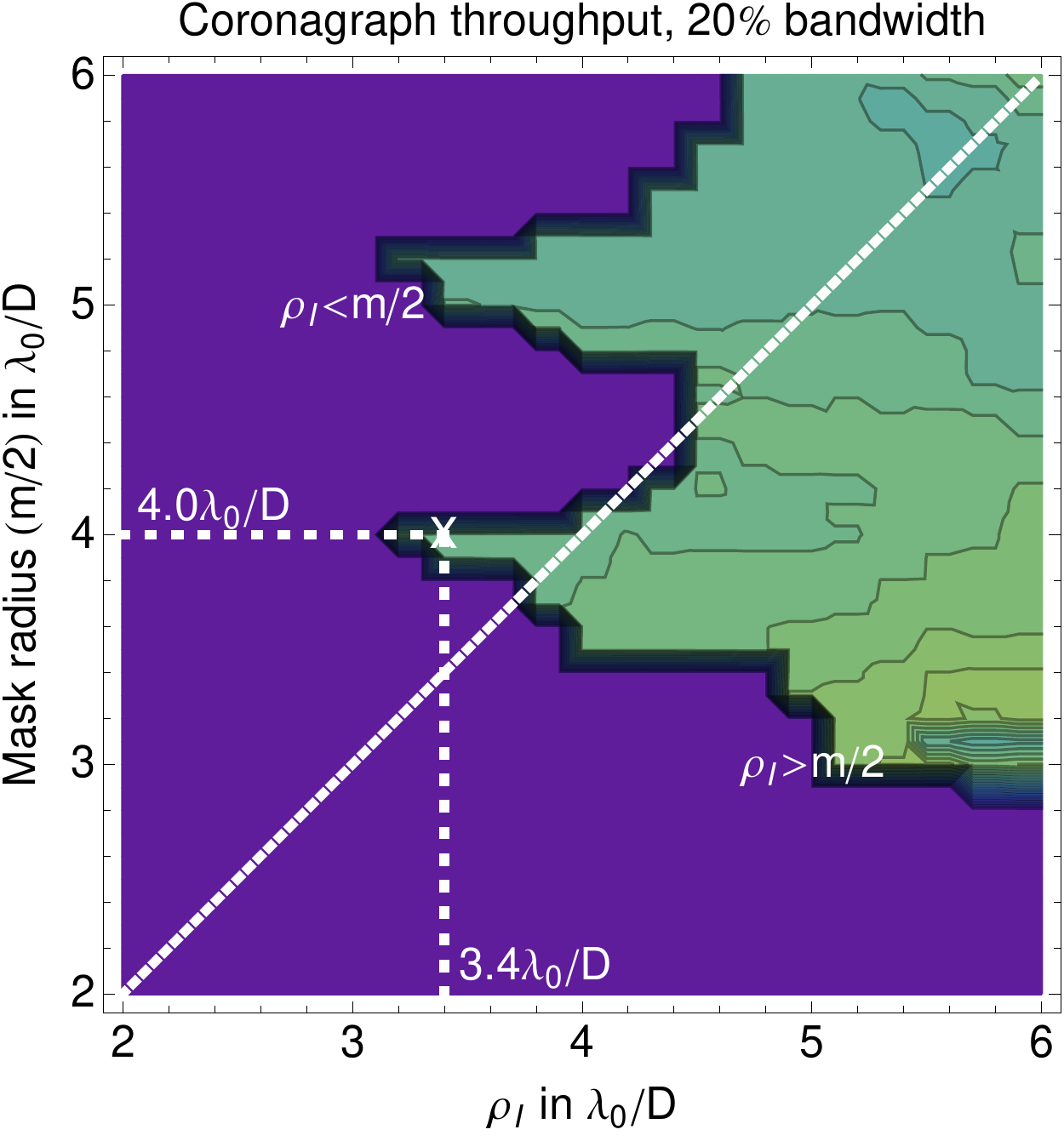}
\includegraphics{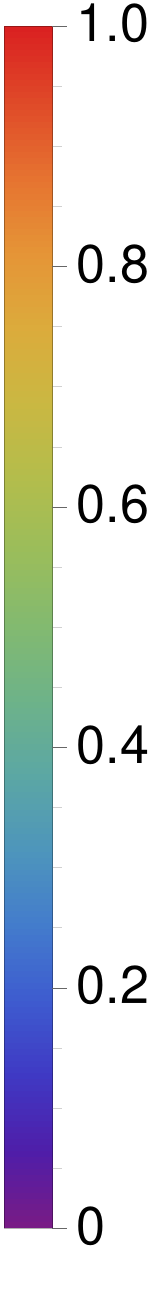}
}
\caption{Coronagraph throughput as a function of the search area inner radius $\rho_i$ and the FPM radius $m/2$ for a circular pupil with 14\% central obscuration in pupil radius ratio (Gemini-like), a 28\% centrally obstructed Lyot stop (ID=2), a search area outer radius $\rho_o = 20\,\lambda_0/D$, and a contrast target $C=8$ for a 20\% bandwidth. Note that one point in this plot corresponds to an APLC design with a value associated with the coronagraph throughput criterion, and that regions with zero throughut (in purple) corresond to cases where the linear pogram does not have a solution. Each point corresponds to a mask radius and an inner edge of the dark region that have been set in the optimization process of the coronagraph. Two different regimes are split by the diagonal white dotted line representing the $\rho_i=m$ case. Surprisingly, solutions are found for $\rho_i < m/2$ (above diagonal), providing coronagraphic images with dark zone below the mask radius. Obviously, this dark region behind the projected mask radius in the final image plane is inaccessible for companion detection and then, no companions are observable at these separations. However, there is room for an enlargement or displacement of the coronagraphic image core due to low-order aberrations without altering the contrast ratio in the dark region above the mask in which planets are expected to be found. These coronagraphic designs are hence expected to be more robust to errors such as telescope jitter in a real system. The cross mark denotes the selected solution for a case where $\rho_i < m/2$ and presented hereafter in Fig. \ref{fig:rho_i_vs_m_solution} with purple curves.}
\label{fig:cp_rhoI_m}
\end{figure}

\begin{figure*}[!ht]
\centering
\resizebox{\hsize}{!}{
\includegraphics{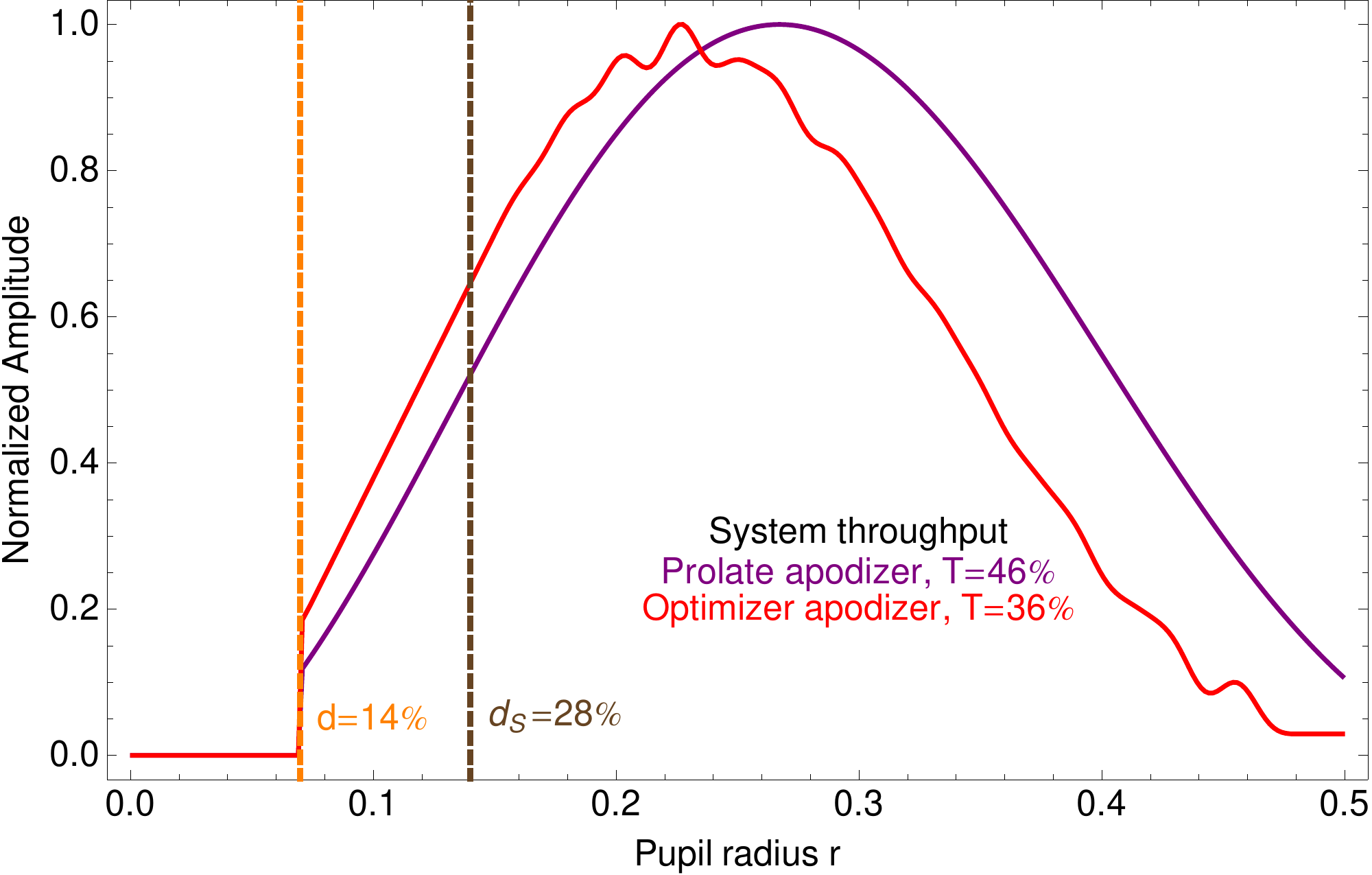}
\includegraphics{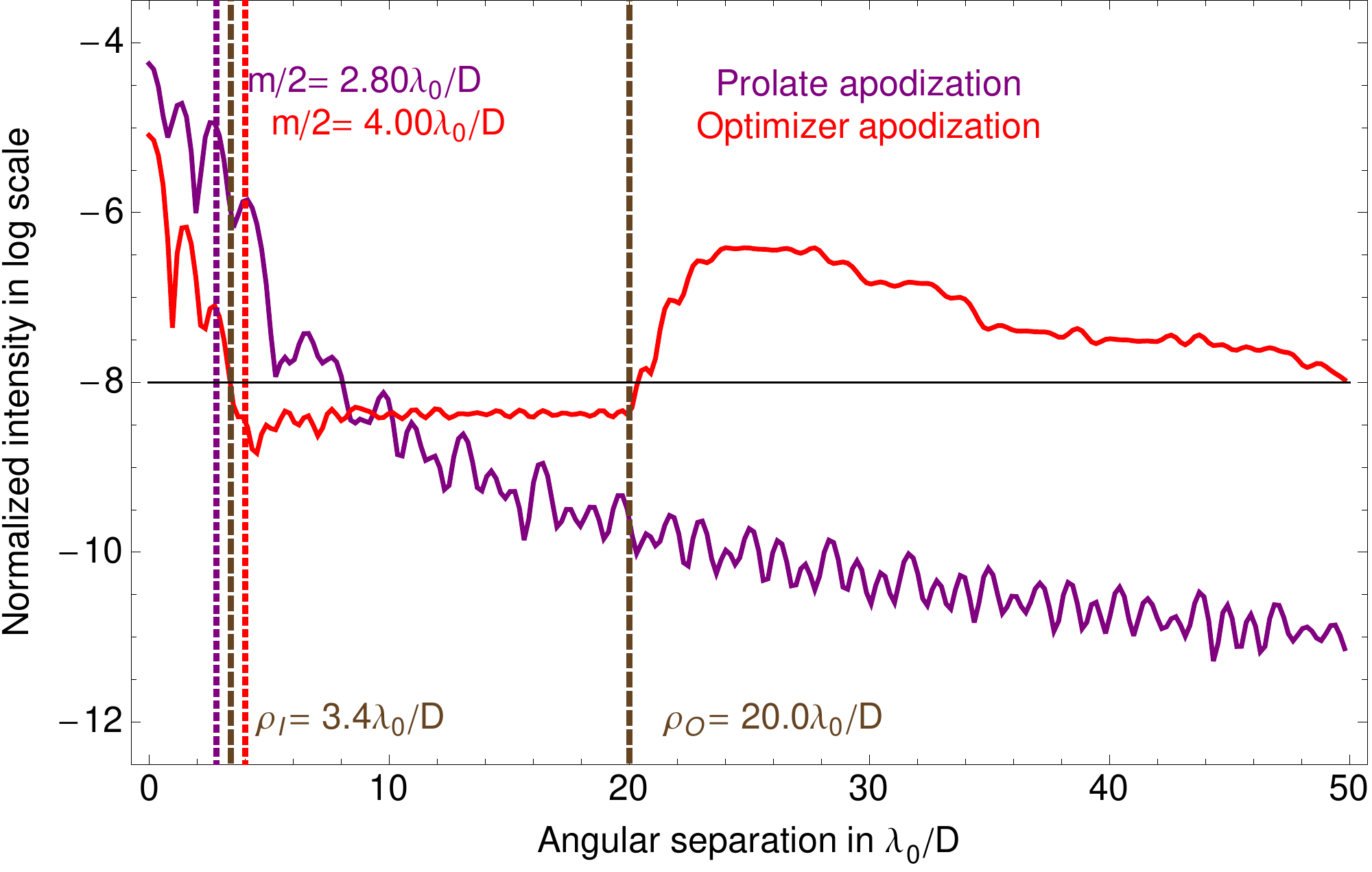}
}
\caption{Illustration of optimized new solutions for the GPI configuration with re-optimization of both the apodizer and FPM, with comparison of APLC designs with apodization using a prolate function and our linear optimization approach. \textbf{Left}: Radial amplitude profiles of the apodizer for 20\% bandwidth, an aperture with 14\% centrally obstruction in pupil radius ratio (orange vertical line) and 28\% centrally obstructed Lyot stop (brown vertical line). Two APLCs are represented here: the GPI design (purple) made with a prolate apodization for a FPM $m/2=2.8\,\lambda_0/D$ as described in \citet{2011ApJ...729..144S} and an optimized design (red) for a FPM radius $m/2=4.0\,\lambda_0/D$, and search area bounds $\rho_i=3.4\,\lambda_0/D$ and $\rho_o=20\,\lambda_0/D$. \textbf{Right}: Radial intensity profiles of the broadband coronagraphic image with these apodizers over the 20\% bandwidth. The focal mask radii, the inner and outer bounds of the search area are delimited with dashed vertical lines. The average broadband contrast is below $10^{-8}$ in the specified dark region for the optimizer design (black horizontal line), showing significant performance improvement over the current GPI design for separations shorter than 8\,$\lambda_0/D$ and enlargement of the discovery space at the shortest distances from the star.}
\label{fig:rho_i_vs_m_solution}
\end{figure*}

Figure \ref{fig:cp_rhoI_m} represents the contrast throughput as a function of the mask radius $m/2$ and inner bound of the dark region $\rho_i$ (see list of parameters in Table \ref{table:rho_i_vs_m}). This plot underlines the existence of coronagraph designs for $\rho_i < m/2$ (see solutions above diagonal), meaning that these solutions generate coronagraphic image dark hole with an inner bound smaller than the mask radius. To illustrate our remarks, we pick up the design denoted by the cross mark in Figure \ref{fig:cp_rhoI_m}. This corresponds to the solution providing the smallest bound for the dark region, see Figure \ref{fig:rho_i_vs_m_solution}.

\begin{table}[!ht]
\caption{Parameters for the panel showed in Figure \ref{fig:cp_rhoI_m}.}
\centering
\begin{tabular}{c c c c}
\hline\hline
Aperture & Bandwidth & Lyot stop & Contrast\\
obstruction (\%) & (\%) & obstruction (\%) & target C\\
\hline
14 & 20 & 28 (ID=2.0) & 8\\
\hline
\end{tabular}\\
\label{table:rho_i_vs_m}
\end{table}

It is important to note that such designs with $\rho_i < m/2$ do not allow to observe companions at distances shorter than $m/2$ since these objects see their light totally blocked by the opaque mask in the intermediate focal plane B of the coronagraph layout. However, there are two important advantages to this configuration:

\begin{itemize}
\item The IWA in the absence of aberrations is solely determined by the geometric IWA from the FPM, and not by the contrast curve of the APLC design, since the dark zone in the PSF extends within the geometrical mask. Therefore this new class of APLC designs decouples completely the IWA from the coronagraph performance itself. The actual ability to observe a faint companion at the shortest separations in this case is therefore entirely determined by the combination of geometric IWA (due to the diameter of the FPM) and wavefront control. 
\item Since the core of the PSF is smaller than the projected angular size of the FPM, this design allows for both expansion or displacement of the PSF core up to the mask radius with no impact on the contrast in the effective dark region. These designs are virtually insensitive to jitter and low-order aberration in this range.
\end{itemize}

A large FPM ($m/2=4\,\lambda_0/D$) is used for the configuration displayed in Figure \ref{fig:rho_i_vs_m_solution}, corresponding to a geometric IWA of $\sim4\,\lambda_0/D$. As a comparison, we recall the GPI configuration for the H-band mask ($m/2=2.8\,\lambda_0/D$), as described in \citet{2011ApJ...729..144S}. Perfect optical surfaces, alignments and absence of atmospheric turbulence are assumed for raw performance comparison between both designs. Performance degradation due to wavefront errors are studied in the following section. Pupil strut effects are not considered in our analysis since they can be mitigated in the Lyot plane with an independant optimization of the Lyot stop \citep{2005ApJ...633..528S,2005ApJ...626L..65S} as it was performed for GPI. Spider vane effects, alignment tolerance and ringing effects from the field stop are considered within such an optimization and diffraction effects from struts hence remain below the $10^{-8}$ raw contrast provided by the optimized circularly symmetric coronagraphs with almost no throughput loss \citep{2009SPIE.7440E..23S,2010SPIE.7735E..86S}. With a larger mask, our design provides a smaller IWA and a higher contrast at small angular separations (between 4\, and 10\,$\lambda_0/D$ or equivalently 170 and 425\,mas in H-band) than the existing coronagraph in GPI.

Even larger masks ($m/2 > 4\,\lambda_0/D$) are worth being studied since the apodizer can be implemented with the Phase-induced Amplitude Apodization \citep[PIAA,][]{2003A&A...404..379G,2003ApJ...599..695T} using a series of 2 aspherical mirrors to reach small angular separations ($<2\,\lambda_0/D$). With such an implementation, the apodizer throughput is unity and the angular resolution units is magnified by the field-independent centroid-based angular magnification defined in \citet{2011JOSAA..28..189P}. With an estimated PIAA magnification of 3.0, our mask radius becomes 1.3\,$\lambda_0/D$ in PIAA units, allowing the observation of planets that are very close to their host star.

In the following, we analyze and compare the tip-tilt sensitivity of several APLCs (including one design optimized with $\rho_i < m/2$).

\section{Low-order aberration sensitivity analysis}\label{sec:sensitivity} 
Low-order aberrations are currently identified as one of the major issues in current and future coronagraphic facilities to reach high-contrast performance and small IWA. Strategies including instrument thermal control, fine low-order wavefront sensing and control are currently investigated to ensure a stable, near-perfect pointing of an observed bright star on the coronagraph in exoplanet direct imagers. Instrumental pointing error constraints can be alleviated by designing tolerant coronagraphs to telescope jitter and focus shift. Support struts also have an impact on coronagraph performance in the presence of low-order aberrations. However, they represent high-spatial frequency structures in the aperture, having a dimmer impact on low-order wavefront errors than the central obscuration and hence, they are not considered in our analysis. Their effects can be mitigated in the Lyot plane with a Lyot stop design that accounts for low-order aberration tolerance.

We here compare the properties of three different APLCs to illustrate the properties of this new type of solutions with GPI as a baseline for comparison and the other two with numerically optimized apodizers with larger FPMs but with the same aperture and Lyot stop geometry (14\% and 28\% central obstruction).

\begin{itemize}
\item the current GPI implementation using a prolate apodization for a mask radius $m/2=2.8\,\lambda_0/D$ in H-band. Made for a 14\% central obstruction and 28\% central obstruction Lyot stop, this design follows the GPI specifications, providing a $10^{-7}$ contrast at 5\,$\lambda_0/D$ (0.2") in H-band \citep{2011ApJ...729..144S}. The design is recalled in Figure \ref{fig:rho_i_vs_m_solution}. 
\item The design identified in Figure \ref{fig:rho_i_vs_ID} and detailed in \ref{fig:GPI_solution}, with the existing GPI mask of radius $m/2 = 3.48\,\lambda_0/D$ in H-band, a dark region inner radius of $\rho_i = 4.0\,\lambda_0/D$, 20\% bandpass, a 14\% central obstruction and 28\% obstructed Lyot stop (ratios are expressed in pupil radius ratio).
\item A second similar design with FPM of radius m/2 = 4.0\,$\lambda_0/D$. The dark region for the design optimization is defined such that the inner bound is smaller the mask radius ($\rho_i = 3.4\,\lambda_0/D$) for a $10^{−8}$ contrast target. The solution is displayed in Figure \ref{fig:rho_i_vs_m_solution} for comparison with the current GPI implementation.
\end{itemize}
The design parameters are listed in Table \ref{table:GPIdesigns}. We now compare the robustness of these three designs to tip tilt errors and higher-order aberrations, and we envision future possible solutions for GPI upgrade.

\begin{table}[!ht]
\caption{Parameters of the three APLCs for sensitivity analysis.}
\centering
\begin{tabular}{c c c c c}
\hline\hline
\multirow{2}{*}{APLC} & \multirow{2}{*}{Apodizer} & Mask radius & Dark region inner & \multirow{2}{*}{Plots}\\
 & & $m/2$ ($\lambda_0/D$) & edge $\rho_i$ ($\lambda_0/D$) & \\
\hline
GPI design in H-band  & prolate  & 2.80 & 5.0 & Fig. \ref{fig:rho_i_vs_m_solution}\\
Optimizer solution 1  & apodizer & 3.48 & 4.0 & Fig. \ref{fig:GPI_solution}\\
Optimizer solution 2  & apodizer & 4.00 & 3.4 & Fig. \ref{fig:rho_i_vs_m_solution}\\
\hline
\end{tabular}\\
\label{table:GPIdesigns}
\tablecomments{All designs are made for a 14\% centrally obstructed aperture and a 28\% obstructed Lyot stop (expressed in pupil radius ratio).}
\end{table}

\subsection{Tip-tilt error and low-order aberrations}
\begin{figure*}[!ht]
\centering
\resizebox{\hsize}{!}{
\includegraphics{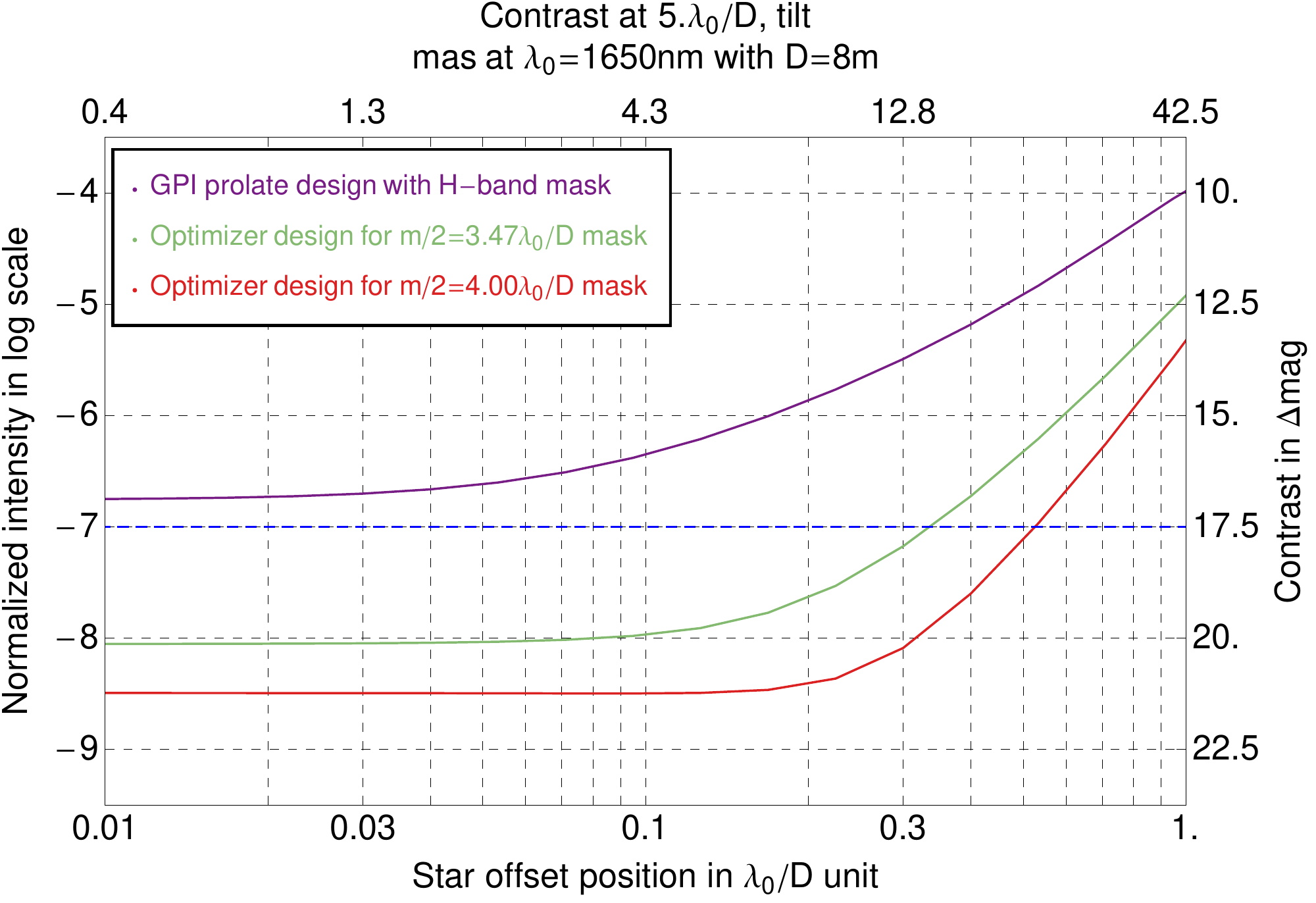}
\includegraphics{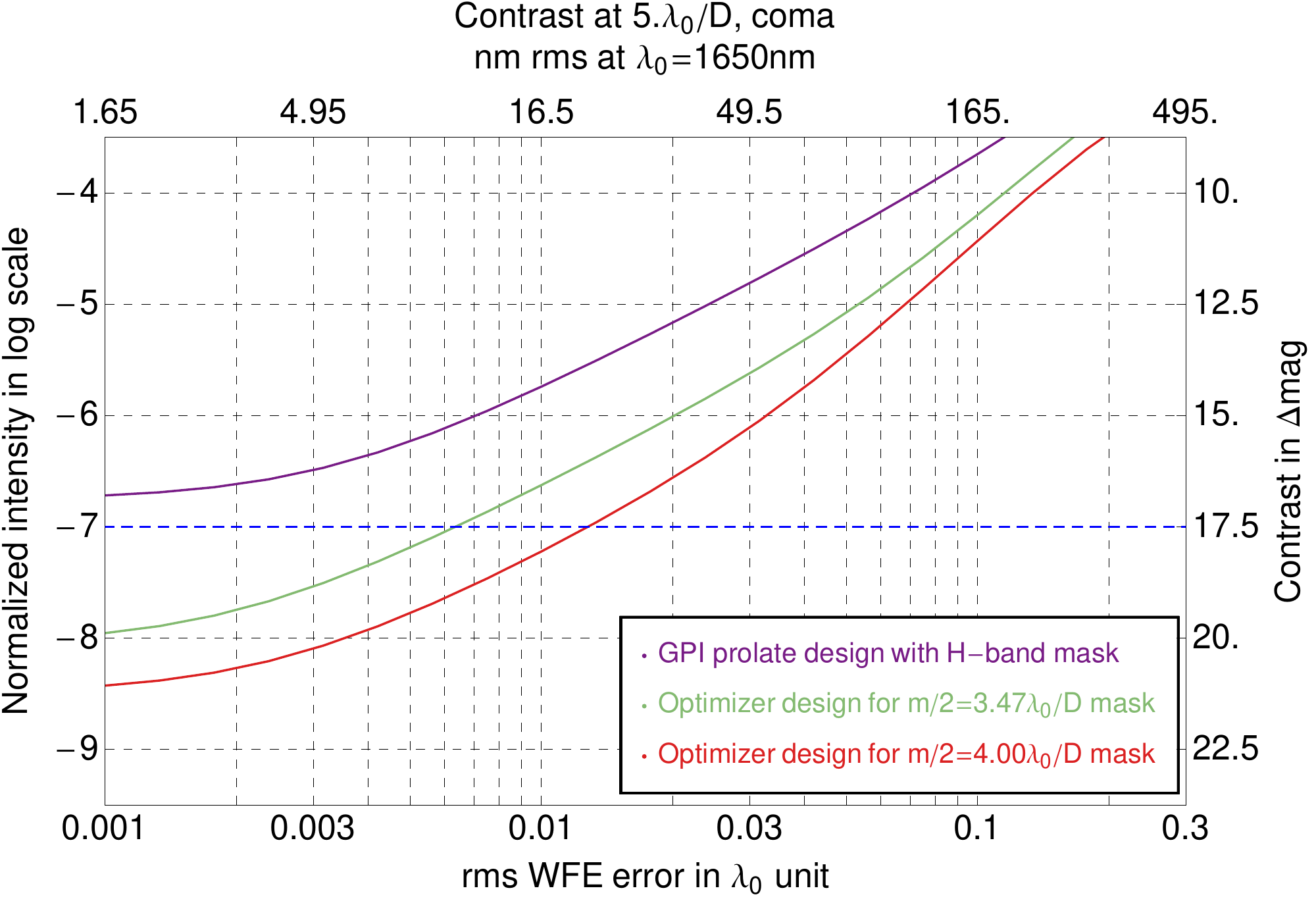}
}
\resizebox{\hsize}{!}{
\includegraphics{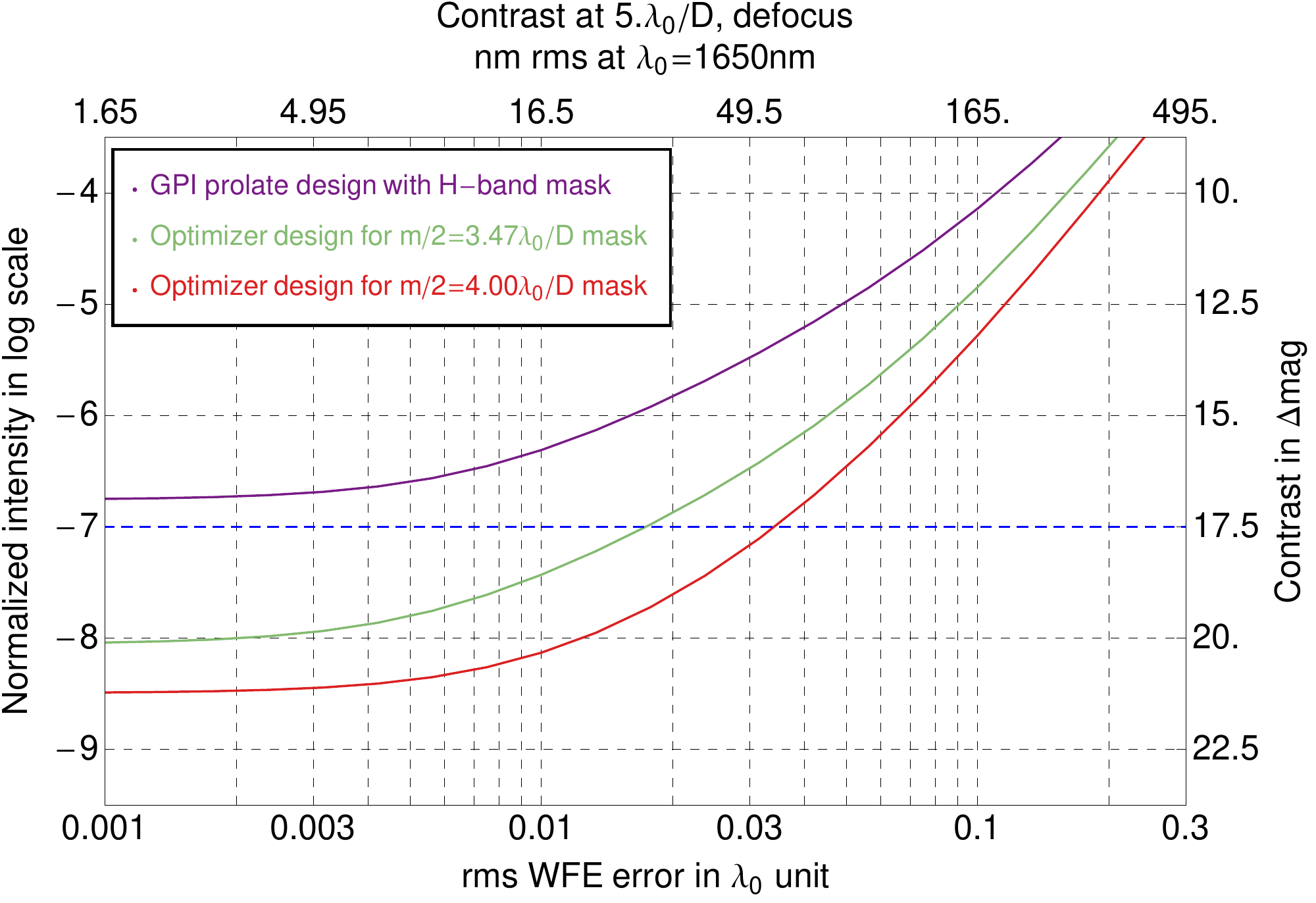}
\includegraphics{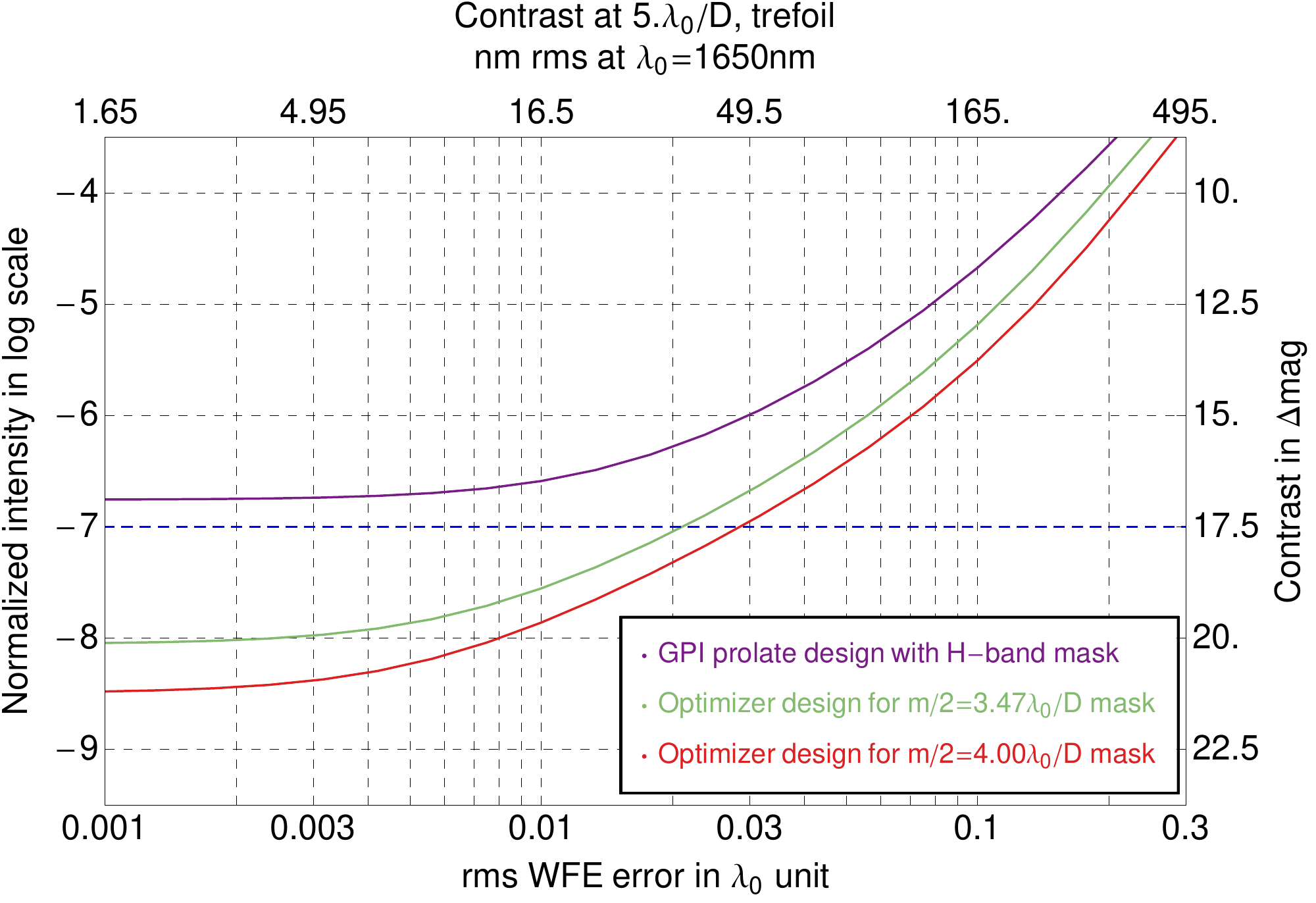}
}
\resizebox{\hsize}{!}{
\includegraphics{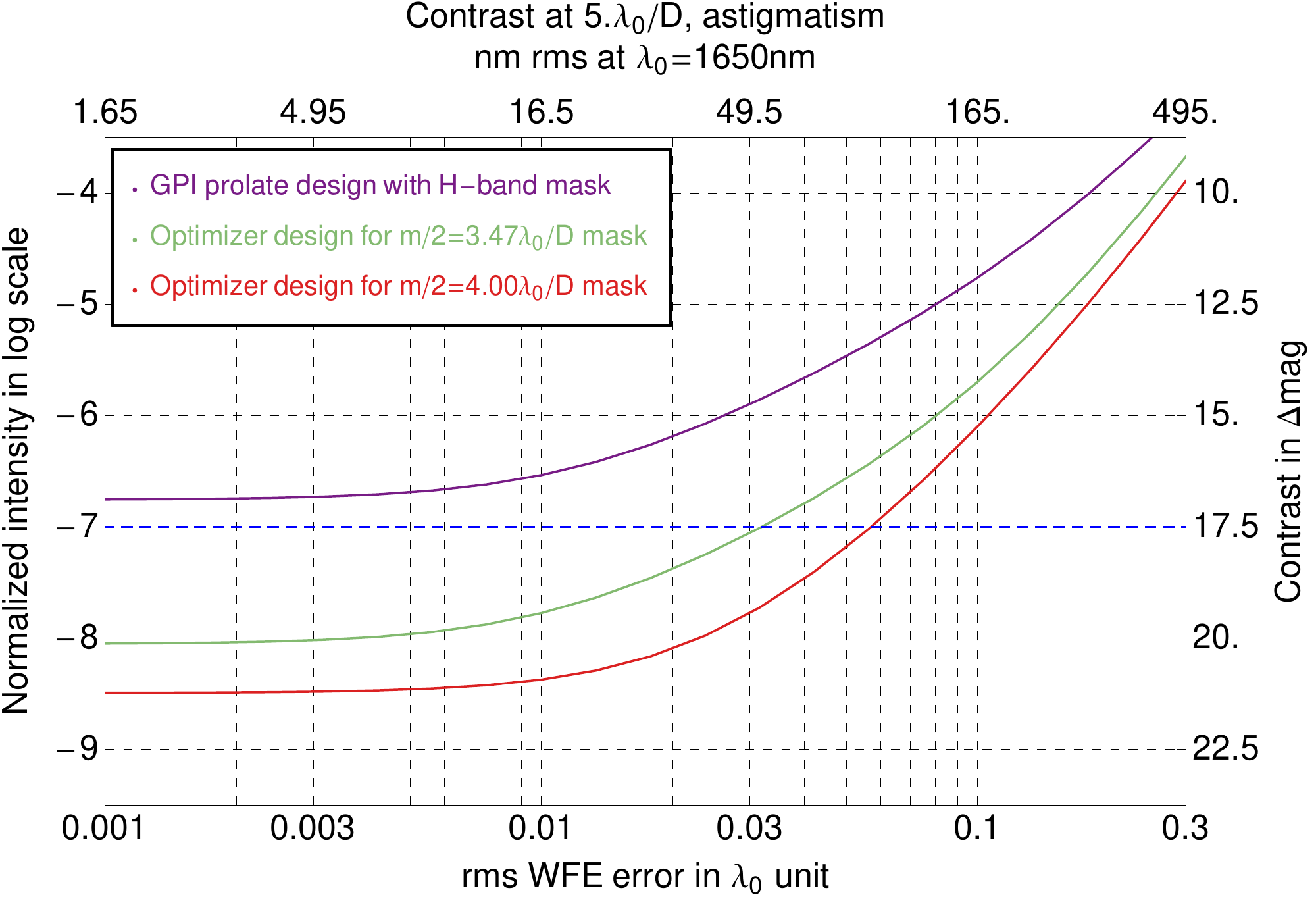}
\includegraphics{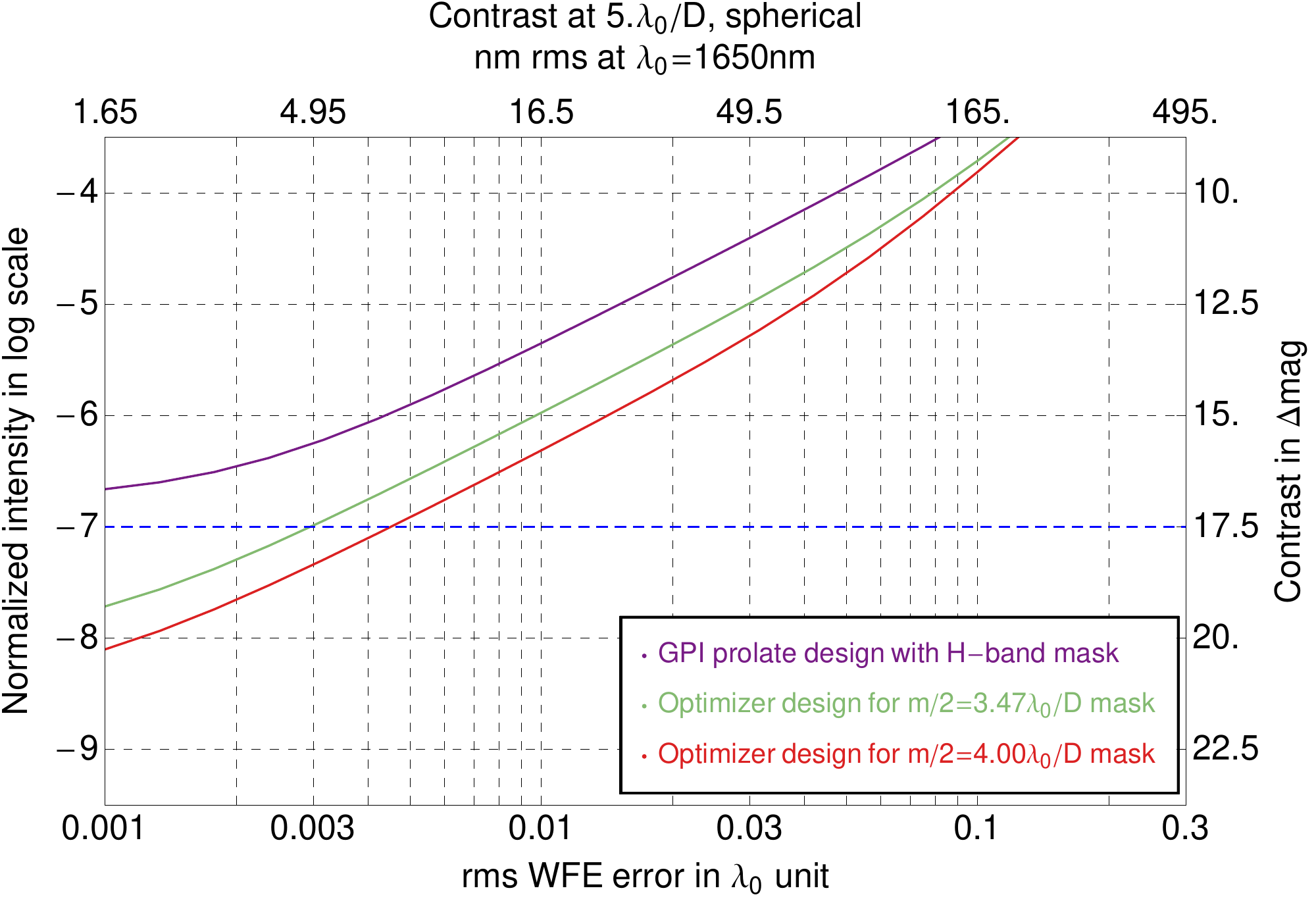}
}
\caption{Average intensity of the broadband coronagraphic image at a 5\,$\lambda_0/D$ angular separation from the optical axis as a function of the wavefront error for a given low-order aberration. From top to bottom and left to right, the six panels correspond to tilt, defocus, astigmatism, coma, trefoil, and spherical aberrations. In each panel, profiles are given for three APLCs working with a 14\% centrally obstructed aperture and 28\% Lyot stop obstruction (ratios are expressed in pupil radius ratio): the GPI design made with a prolate apodization for a FPM $m/2=2.8\,\lambda_0/D$ as described in \citet{2011ApJ...729..144S} and two designs with different couples of mask size and search area inner bound: \{design 1, $m/2=3.48\,\lambda_0/D$; $\rho_i=4.0\,\lambda_0/D$\} and \{design 2, $m/2=4.0\,\lambda_0/D$; $\rho_i=3.4\,\lambda_0/D$\}, see parameters in Table \ref{table:GPIdesigns}. The blue solid line denote the $10^{-7}$ intensity levels specified for GPI coronagraphs. For each type of aberration, our coronagraphs maintain an intensity level below $10^{-7}$ at larger wavefront errors than the GPI design, showing low-order aberration sensitivity improvement. For example in the case of tilt (top left panel), our optimizer design 2 curve presents a plateau and intensity levels below $10^{-8}$ for pointing errors up to 0.25\,$\lambda_0/D$ (11\,mas for an 8\,m telescope in H-band), underlining the contrast performance stability of our coronagraph design compared to the GPI implementation.}
\label{fig:lo_sensitivity}
\end{figure*}

Figure \ref{fig:lo_sensitivity} shows the contrast performance at 5\,$\lambda_0/D$ of the three APLCs as a function of tip tilt, defocus, astigmatism, coma, trefoil and spherical aberration in each of the 6 panels. In the tilt case (top left panel), the curves of our optimizer designs present a plateau for pointing errors ranging from 0 to 0.25\,$\lambda_0/D$ (11\,mas for an 8\,m class telescope in H-band) with an intensity level below $10^{-7}$ for tip-tilt smaller than $0.5\,\lambda_0/D$ (22\,mas) while the profile for the current GPI design exhibits a contrast loss as the star pointing error increases with an intensity level above $10^{-7}$. For the other aberrations, our designs also show an intensity level is one to two order of magnitude smaller than the current GPI design.

Our designs show a clear improvement both in contrast and in robustness to low-order aberrations compared with the GPI current implementation. This is particularly interesting in the context of GPI for which telescope jitter has been identified as an important issue in the first runs of the instrument \citep[beyond the 3 mas initially specified during the instrument design process,][]{2014arXiv1407.7893H}. More generally, jitter and vibrations are a potential concern for any coronagraphic instrument either on the ground or in space, and optimizing instrument designs with increased robustness to tip/tilt errors is therefore highly valuable. 

Our new designs with the optimizer remove the diffraction at smaller angular separations from the star than with the current GPI design (i.e. between 4\, and 5\,$\lambda_0/D$), and increase the raw contrast between 4\, and 8\,$\lambda_0/D$ by one order of magnitude or more. However the coronagraph throughput is reduced slightly from 46\% with GPI implementation to 36\% in our best case.

Our designs constitute promising options for a potential coronagraph upgrade for GPI. In particular, the insertion of the design with $m/2=3.48\,\lambda_0/D$ only requires the replacement of the apodization in the filter wheels of the instrument. The insertion of the design with $m/2=4.0\,\lambda_0/D$ requires the introduction of a new FPM in addition to replacement of an apodizer, offering an even larger discovery space and a more robust system to low-order aberrations than the design mentioned above.

These results are detailed for the GPI design as an illustration comparison point without loss of generality, and apply to any other coronagraphic configuration. 

\subsection{stellar angular size}
The study of the robustness to tip-tilt errors can be extended to stellar angular sizes. Indeed, an on-axis resolved star can be seen as a combination of off-axis point sources at all the angular positions ranging up to the stellar angular radius. This is equivalent to consider a sum of tip-tilt point sources weighted with their offset position. We can devise the coronagraph contrast sensitivity at 5\,$\lambda_0/D$ by averaging the solid-line profiles showed in Figure \ref{fig:lo_sensitivity} with weights corresponding to the offset positions. 

\begin{figure}[!ht]
\centering
\resizebox{\hsize}{!}{
\includegraphics{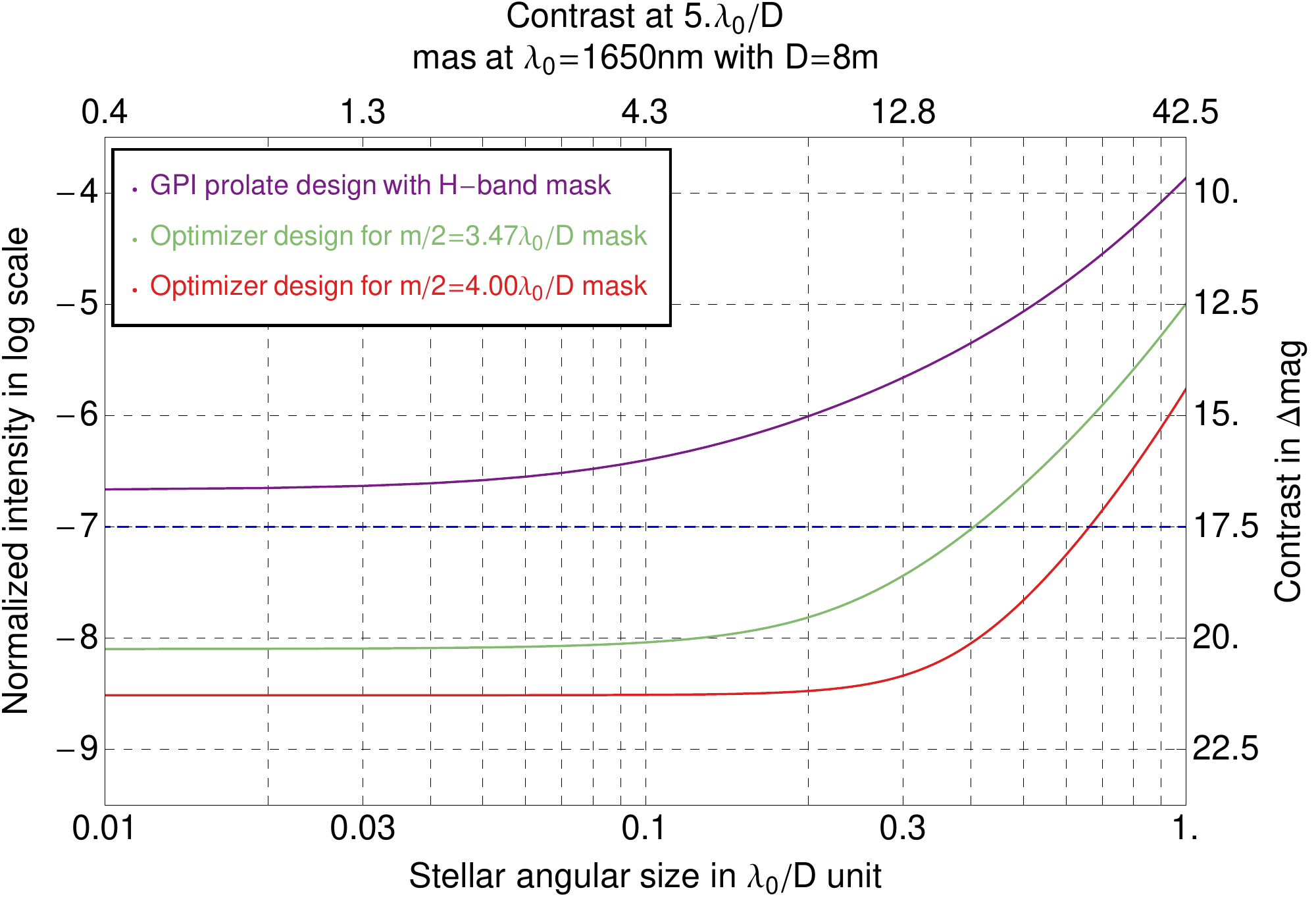}
}
\caption{Average intensity of the broadband coronagraphic image at a 5\,$\lambda_0/D$ angular separation from the optical axis as a function of the stellar angular size. Profiles are given for three APLCs working with a 14\% centrally obstructed aperture and 28\% Lyot stop obstruction (ratios are expressed in pupil radius ratio): the GPI design made with a prolate apodization for a FPM $m/2=2.8\,\lambda_0/D$ as described in \citet{2011ApJ...729..144S} and two designs with different couples of mask size and search area inner bound: \{design 1, $m/2=3.48\,\lambda_0/D$; $\rho_i=4.0\,\lambda_0/D$\} and \{design 2, $m/2=4.0\,\lambda_0/D$; $\rho_i=3.4\,\lambda_0/D$\}, see parameters in Table \ref{table:GPIdesigns}. Blue solid line denotes the $10^{-7}$ intensity levels specified for the GPI coronagraphs. Our designs provide a contrast performance with an order of magnitude better than GPI design. In addition, our optimizer design 2 curve presents a plateau and intensity levels below $10^{-8}$ for stellar angular size up to 0.25\,$\lambda_0/D$ (11\,mas for an 8\,m telescope in H-band), underlining the contrast performance stability of our coronagraph design compared to the GPI implementation.}
\label{fig:star_size}
\end{figure}

Figure \ref{fig:star_size} displays the contrast performance at 5\,$\lambda_0/D$ of the three designs as a function the angular size of the observed objects (dashed curves). As with the pointing errors, our optimizer designs are more robust to stellar angular size than the current GPI implementation. 

\section{Conclusion}
We have presented novel APLC designs that are specifically optimized to meet a given level of contrast in a dark region in the broadband coronagraphic PSF. These solutions rely on the linear relation between the coronagraphic electric field and the entrance pupil apodization at any wavelength, and on Linear Programming optimization similar to those developed for Shaped Pupil concepts \citep[e.g][]{2003ApJ...582.1147K,2003ApJ...599..686V,2011OExpr..1926796C}.

We developed solutions for Gemini/GPI to propose future possible upgrades. Within a dark region ranging between 5\,$\lambda_0/D$ and 20\,$\lambda_0/D$ and over a 20\% bandwidth and in the presence of a 14\% central obstruction our solution offers a $10^{−8}$ contrast, which is compatible with the contrast specifications for the instrument. 

The major advance of the proposed solutions is the ability to produce dark holes with an inner bound smaller than the FPM radius. Such concepts are extremely interesting for their very high tolerance to low-order aberrations including jitter and focus shift, and the proposed designs are virtually insensitive to jitter or tip tilt up to $\pm$10\,mas.

In this context, we analyzed our solutions and compared them with the current GPI implementation in H- band in terms of contrast performance and robustness to low-order aberrations. Our new designs increase considerably the robustness to low-order aberrations by a factor of several orders of magnitude theoretical contrast gain for a system throughput higher than 35\% (about a 10\% reduction compared to GPI). These results are encouraging for future potential upgrades of APLCs on GPI, SPHERE and P1640.

To a larger extent, our new approach allows us to propose APLCs with high-contrast performance and robustness to low-order aberrations. This is particularly interesting in the context of future space missions (WFIRST-AFTA [\citealt{2013arXiv1305.5422S}], EXO-C, ATLAST [\citealt{2012OptEn..51a1007P}], etc) for which controlling low order aberrations and jitter is an important challenge. Robust coronagraphs are complementary to low-order wavefront sensing and control to reach stable high-contrast performance \citep{2013A&A...555A..94N}. The APLC type coronagraph therefore constitutes a promising option for future exoplanet direct imaging missions. APLC can be combined with PIAA as the apodization implementation to reach smaller IWA \citep[e.g. PIAACMC,][]{2010ApJS..190..220G}.

Our APLCs were designed with one-dimensional apodization to consider the case of axisymmetric pupils (telescope circular aperture with large central obstruction) and can be used in the presence of aperture with struts by mitigating the diffraction effects in the Lyot plane with an optimized Lyot stop. They can also be combined with the Active Compensation of Aperture Discontinuities \citep[ACAD,][]{2013ApJ...769..102P}, a method based on the use of two deformable mirror to mitigate the diffraction effects caused by the pupil spiders. This promising technique can alleviate some constraints on future coronagraph designs.

Two-dimensional apodizations for complex aperture geometries are also currently being investigated with an hybrid combination of APLC and shaped pupil to address the case of aperture with struts (spiders, pupil segmentation, etc). This study builds on the same approach (numerical optimization of APLCs, combined with a two-dimensional optimization of shaped pupil apodizers \citep{2011OExpr..1926796C}. 

Our multi-wavelength approach was applied to APLCs but we are currently investigating its extension to phase mask coronagraphs, such as the vector vortex \citep{2005ApJ...633.1191M,2013ApJS..209....7M} or the dual-zone phase mask concepts \citep{2003A&A...403..369S,2012A&A...538A..55N}, to develop small inner working angle solutions.

\acknowledgments
This work is supported by the National Aeronautics and Space Administration under Grants NNX12AG05G and NNX14AD33G issued through the Astrophysics Research and Analysis (APRA) program (PI: R. Soummer). This material is also partially based upon work carried out under subcontract 1496556 with the Jet Propulsion Laboratory funded by NASA and administered by the California Institute of Technology. The authors would like to thank Marshall D. Perrin and Alexis Carlotti for very helpful discussions, and Colin Norman for stimulating exchanges on mathematical aspects at the early stage of the manuscript.

\section{Appendix}
\subsection{$10^{-10}$ broadband contrast with APLC}
\begin{figure*}[!ht]
\centering
\resizebox{\hsize}{!}{
\includegraphics[height=5.0cm]{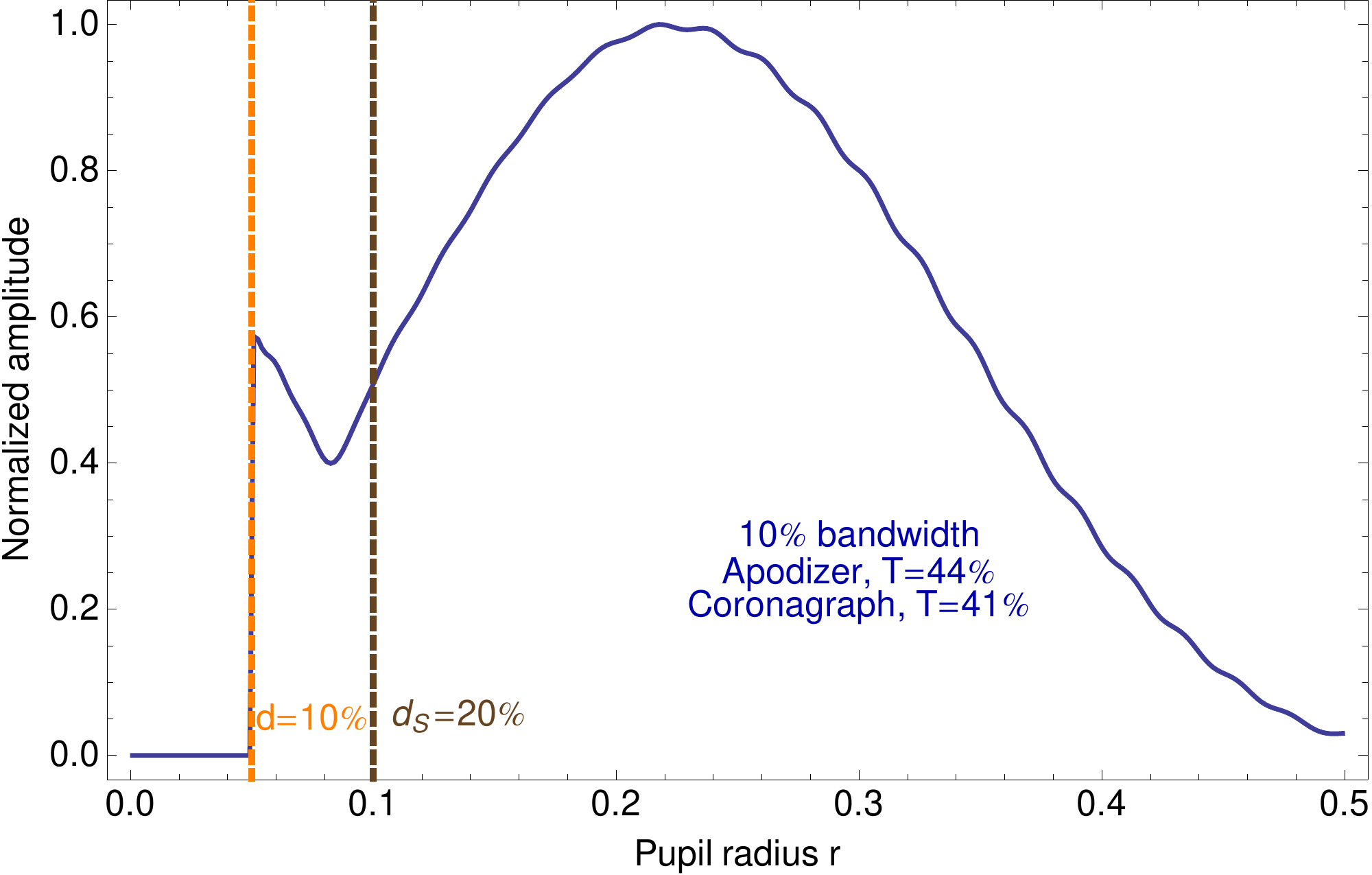}
\includegraphics[height=5.0cm]{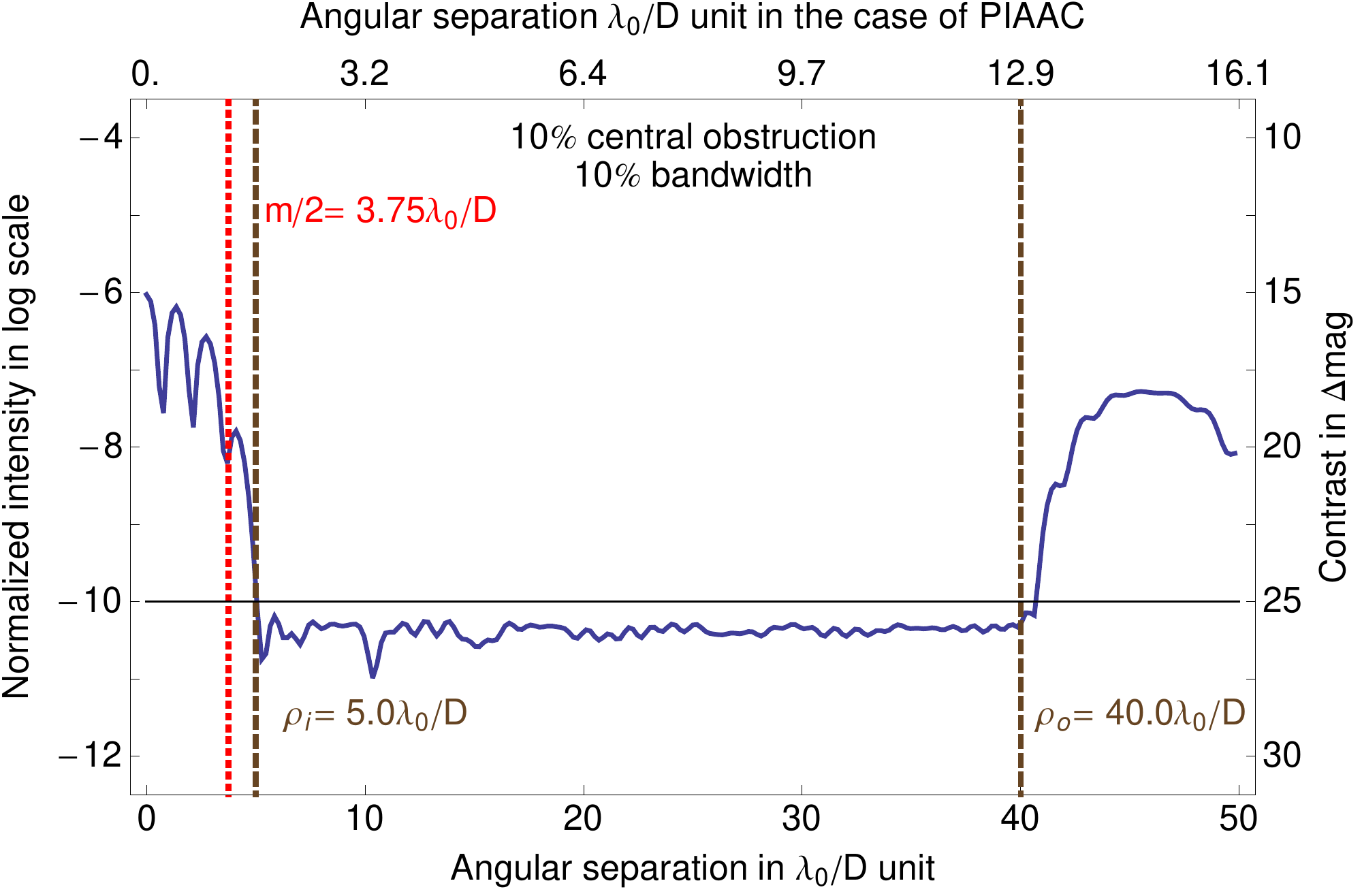}
}\\
\resizebox{\hsize}{!}{
\includegraphics[height=5.0cm]{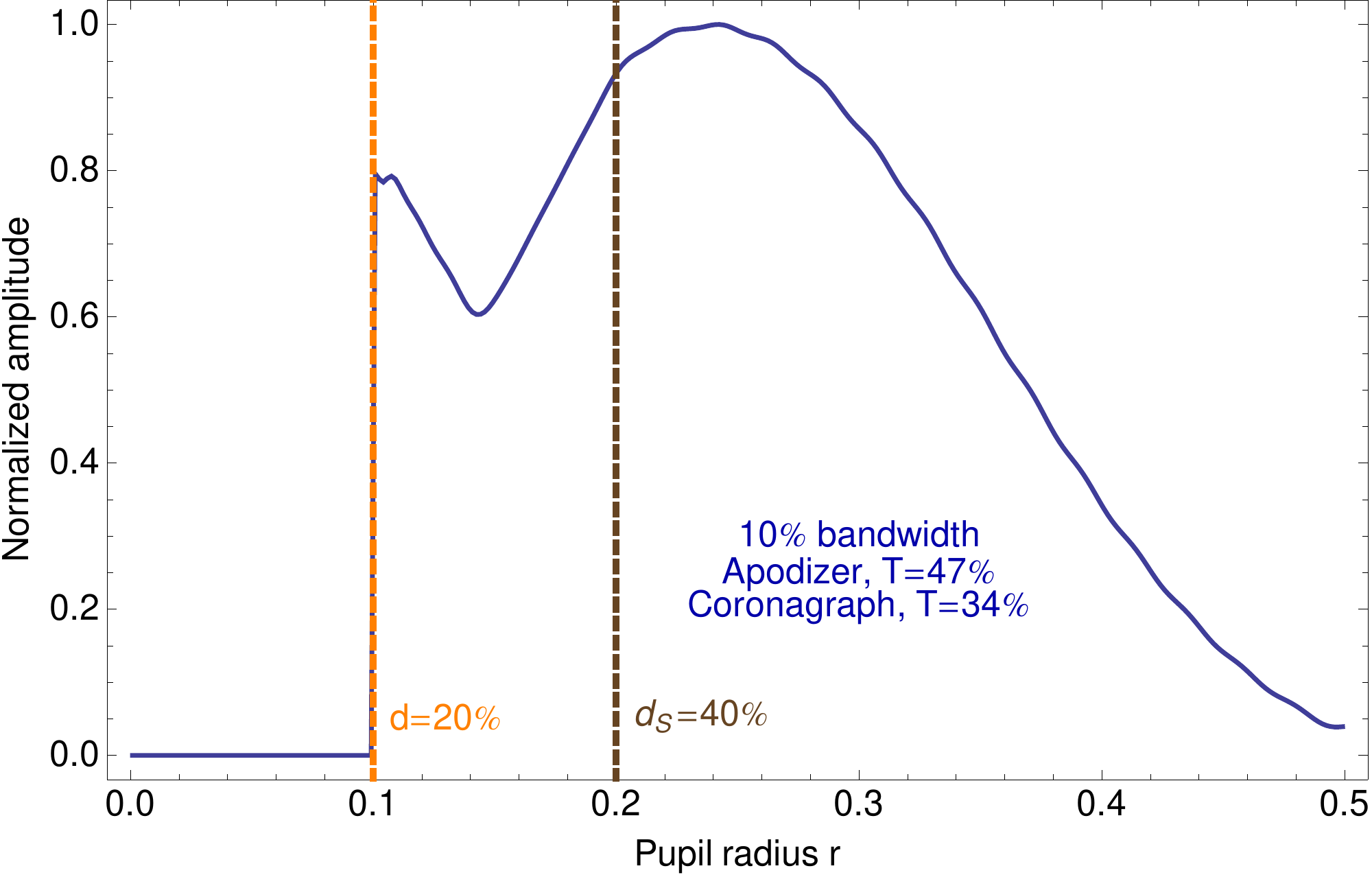}
\includegraphics[height=5.0cm]{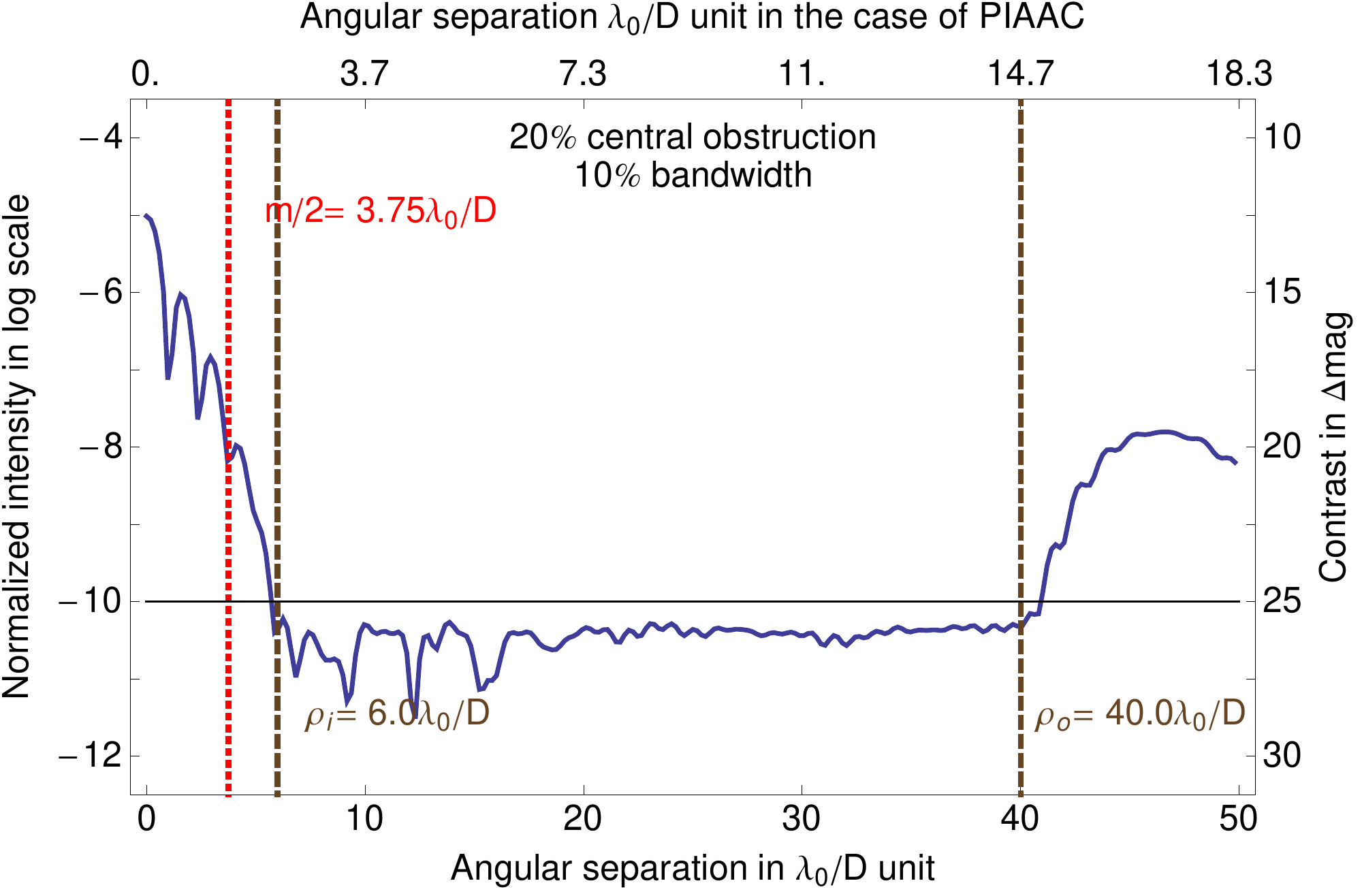}
}\\
\resizebox{\hsize}{!}{
\includegraphics[height=5.0cm]{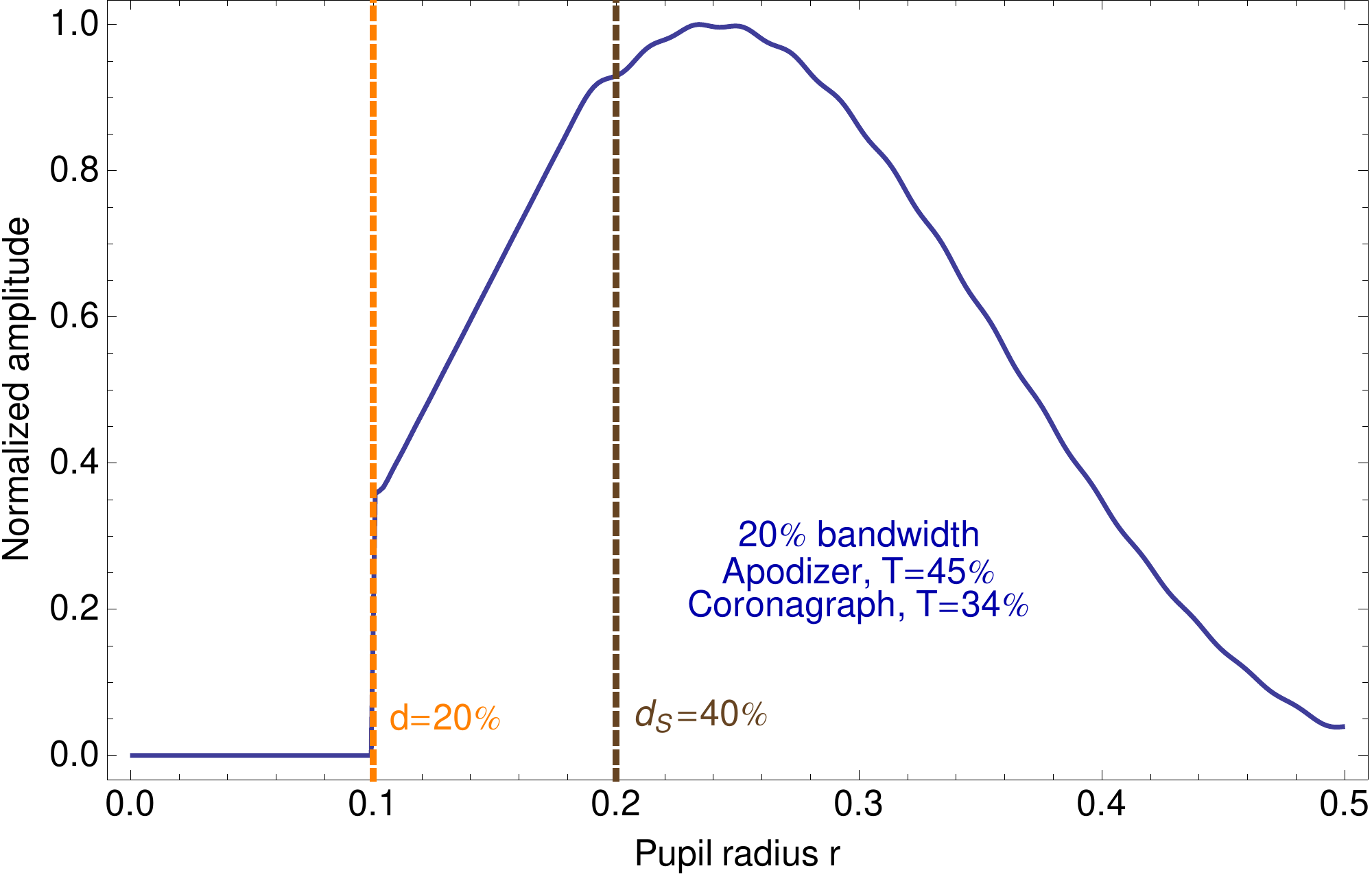}
\includegraphics[height=5.0cm]{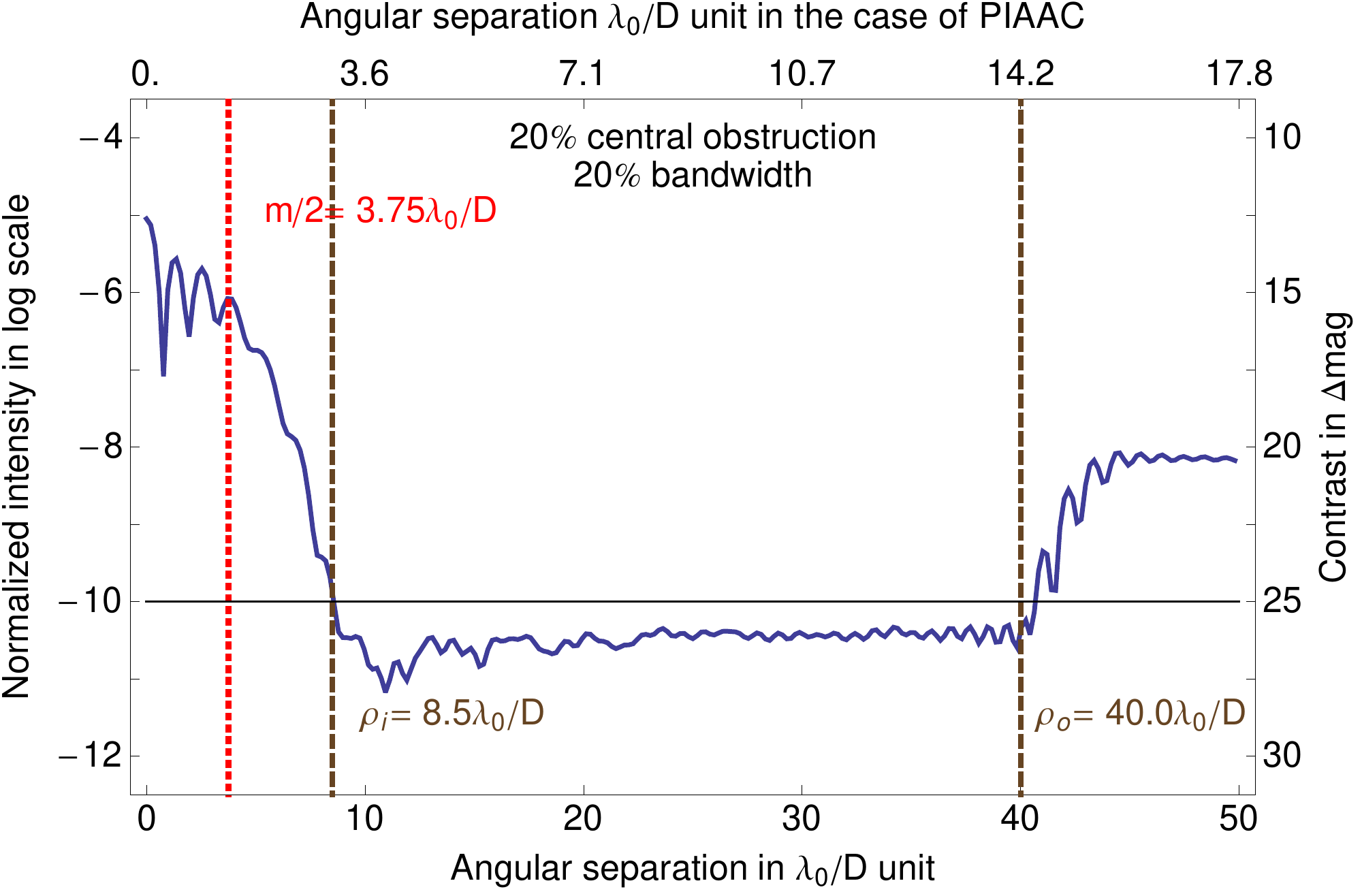}
}
\caption{Optimal APLC designs for a $10^{-10}$ broadband contrast in the search area with three combinations of central obstruction and bandwidth (from top to bottom): (10\%, 10\%), (20\%, 10\%), and (20\%, 20\%). \textbf{Left}: radial amplitude profiles of the apodization. The orange and brown lines delimit the obscuration radius of the entrance pupil and Lyot stop. We set the Lyot stop obstruction oversizing factor to 2 and the apodizer derivative limit to $b=5$ for all the configurations. \textbf{Right}: Coronagraphic intensity profiles reached by APLC in broadband light with the apodizers shown on the left panel. We present our PSFs in two possible configurations: when the apodization is achieved using a gray-scale screen (APLC; bottom x-axis for \ldq on-sky\rdq $\lambda_0/D$), the \ldq on-sky\rdq $\lambda_0/D$ bottom x-axis, and when the apodization is achieved via two pupil remapping mirrors (PIAA; top x-axis for \ldq on-sky\rdq $\lambda_0/D$), the \ldq on-sky\rdq $\lambda_0/D$ top x-axis. The blue dashed lines represent the bounds of the search area $\mathcal{D} $ ($\rho_o=40\,\lambda_0/D$ and from top to bottom, $\rho_i=$5, 6, and 8.5\,$\lambda_0/D$). The red dot line delimits the FPM radius, set to $m/2=3.75\,\lambda_0/D$ for all the configurations. As we enlarge the central obstruction or the spectral bandwidth, the inner bound of the search area $\rho_i$ is increased to reach the contrast target while obtaining a high-throughput apodizer. Solutions are also found for a search area with a smaller inner bound $\rho_i$ by relaxing the constraint on the apodizer derivative, but at the cost of smaller throughput for the apodization.}
\label{fig:1e10contrast}
\end{figure*}

Our studies have been mainly focused on designing APLC for an aperture with 14\% centrally obstruction in pupil radius ratio to yield a $10^{-8}$ contrast in the search area $\mathcal{D}$, offering promising possible upgrades for GPI and SPHERE, and for the baseline design of the HiCAT testbed coronagraph \citep{2013SPIE.8864E..1KN,2014arXiv1407.0980N}. We here provide some examples of APLCs for different central obstruction sizes (10\% and 20\%) to produce $10^{-10}$ broadband PSFs dark region, see Figure \ref{fig:1e10contrast}. These designs produces a large dark region inner radius ($> 5\,\lambda_0/D$) but they can be combined with PIAA as the apodization implementation to reach small IWA \citep[e.g. PIAACMC,][]{2010ApJS..190..220G}, proving promising options for future exoplanet direct imaging missions.

\subsection{Solutions for very large obstructed aperture}
\begin{figure*}[!ht]
\centering
\resizebox{\hsize}{!}{
\includegraphics[height=5.0cm]{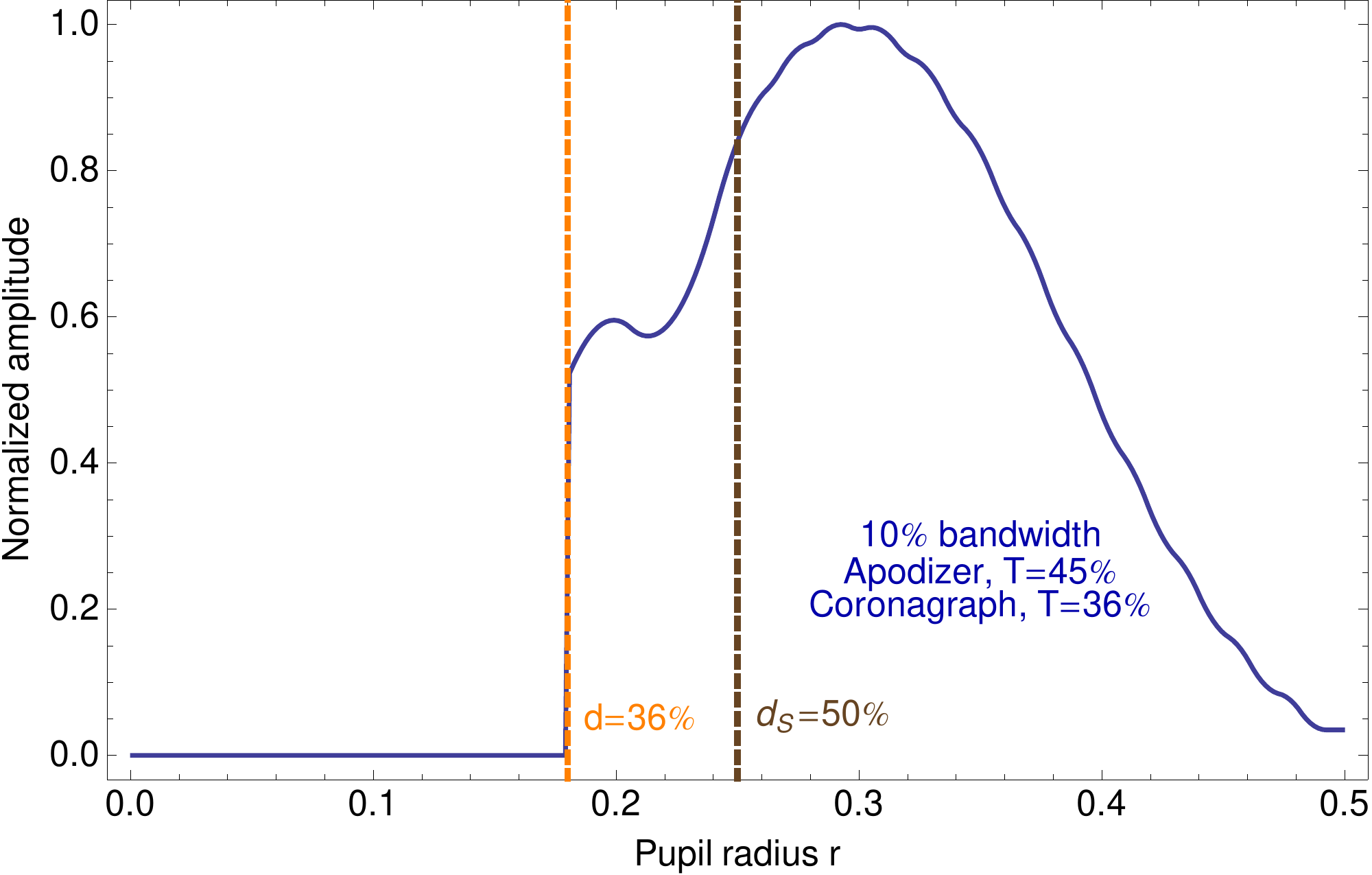}
\includegraphics[height=5.0cm]{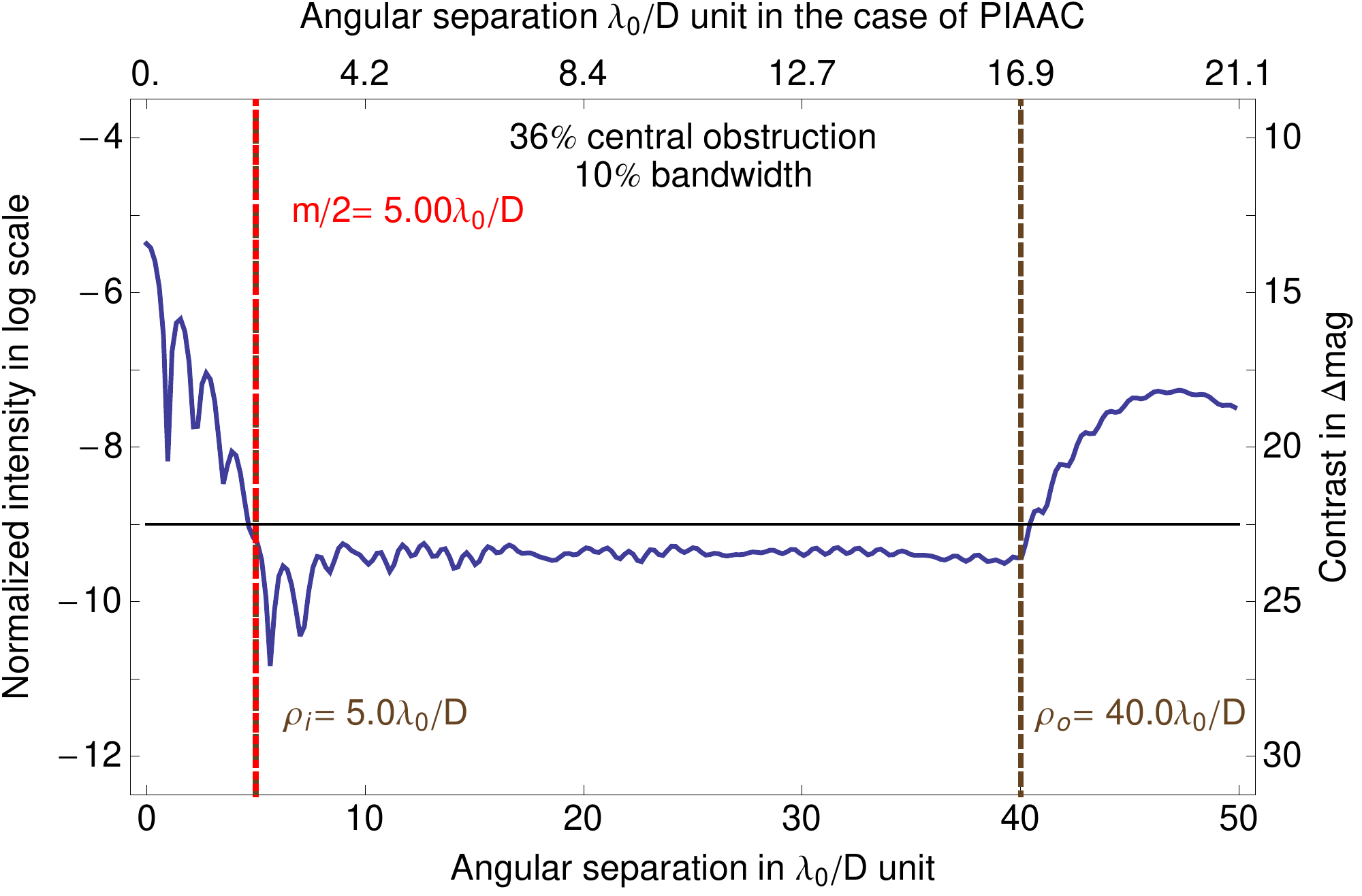}
}
\caption{Example of APLC design for a very large central obstruction (36\%) with a $10^{-9}$ contrast in the search area over 10\% bandwidth. \textbf{Left}: radial amplitude profiles of the apodization. The orange and brown lines delimit the obscuration radius of the entrance pupil and Lyot stop. We set the Lyot stop obstruction with an ID oversizing factor to 1.4. \textbf{Right}: Coronagraphic intensity profile reached by APLC in broadband light with the apodizer shown on the left panel. We present our PSFs in two possible configurations: when the apodization is achieved using a gray-scale screen (APLC; bottom x-axis for \ldq on-sky\rdq $\lambda_0/D$), the \ldq on-sky\rdq $\lambda_0/D$ bottom x-axis, and when the apodization is achieved via two pupil remapping mirrors (PIAA; top x-axis for \ldq on-sky\rdq $\lambda_0/D$), the \ldq on-sky\rdq $\lambda_0/D$ top x-axis. The blue dashed lines represent the bounds of the search area $\mathcal{D}$ ($\rho_I=5\,\lambda_0/D$ and $\rho_o=40\,\lambda_0/D$). The red dot line delimits the FPM radius, set to $m/2=5.0\,\lambda/D$.}
\label{fig:afta}
\end{figure*}

Our previous solutions have been given for apertures with obstructions up to 20\%. However, telescopes such as WFIRST-AFTA in space or ELTs on the ground will have aperture with very large central obstructions ($>30\%$). We here give an example of solutions for such geometry with a design for 36\% central obstruction, see Figure \ref{fig:afta}. This APLC design produces star images with $10^{-9}$ contrast dark region for a 20\% bandpass. Such a solution can also be combined with a PIAA as an apodization implementation to reach small IWA (down to $2\,\lambda_0/D$).

\bibliographystyle{apj}   
\bibliography{APLC_IV_v09_arxiv}   

\end{document}